\documentclass[reprint,amsmath,amssymb,aps]{revtex4-1}
\usepackage{graphicx}
\usepackage{dcolumn}
\usepackage{bm}
\usepackage{float}
\usepackage{color}
\usepackage[dvipsnames]{xcolor}
\usepackage{hyperref}
\usepackage{graphicx}
\usepackage{braket}
\usepackage{appendix}
\usepackage{chemmacros}
\hypersetup{colorlinks=true, citecolor=blue, urlcolor=blue, linkcolor=blue}

\begin{document}

\preprint{APS}

\title{Exploration of trivial and non-trivial electronic phases and of collinear and non-collinear magnetic phases in low-spin d$^5$ perovskites}
\author{Amit Chauhan}
\author{B. R. K. Nanda}
 \email{nandab@iitm.ac.in}
 
\affiliation{1. Condensed Matter Theory and Computational Lab, Department of Physics,
Indian Institute of Technology Madras, Chennai - 600036, India\\
}
\affiliation{
2. Center for Atomistic Modelling and Materials Design,
Indian Institute of Technology Madras, Chennai - 600036, India\\
}
\affiliation{
3. Functional Oxide Research Group,
Indian Institute of Technology Madras, Chennai - 600036, India
}

\date{\today}

\begin{abstract}
The $4d$ and $5d$ transition metal oxides have become important members of the emerging quantum materials family due to competition between onsite Coulomb repulsion ($U$) and spin-orbit coupling (SOC). Specifically, the systems with $d^5$ electronic configuration in an octahedral environment are found to be capable of posessing invariant semimetallic state and perturbations can lead to diverse magnetic phases. In this work, by formulating a multi-band Hubbard model and performing SOC tunable DFT+$U$ calculations on a prototype SrIrO$_3$ and extending the analysis to other iso-structural and isovalent compounds, we present eight possible electronic and magnetic configurations in the $U$-SOC phase diagram that can be observed in the family of low-spin $d^5$ perovskites. They include the protected Dirac semimetal state, metal and insulator regimes, collinear and noncollinear spin ordering. The latter is explained through connecting hopping interactions to the rotation and tilting of the octahedra as observed in GdFeO$_3$. Presence of several soft phase boundaries makes the family of $d^5$ perovskites an ideal platform to study electronic and magnetic phase transitions under external stimuli.
\end{abstract}

\maketitle

\section{Introduction}
In the past decade the discovery that spin-orbit coupling leads to the formation of symmetry protected conducting surface or edge states in a wide range of topological class of compounds, has attracted a great deal of attention among the researchers working in quantum many-body theory \cite{Hasan2010,Qi2011}. The SOC is a relativistic effect and it is treated as a small perturbation in solids. But for systems with heavy elements, it varies as fourth power of the atomic number and therefore it has an intriguing influence on the electronic and magnetic structure of such systems. In the family of correlated oxide systems, where the onsite Coulomb repulsion with other competing interactions such as Hund's coupling ($J_\mathrm{H}$), crystal field ($\Delta$) and electron hopping strength ($t$) create a rich electronic and magnetic phase diagram, the presence of SOC bridges the symmetry protected topologically invariant states and magnetism. Moving down the periodic table from $3d$ to $4d$ to $5d$ elements, while the SOC increases, the $d$ orbitals become more spatially extended to reduce the correlation strength $U$. For the $5d$ families of oxides, both SOC and $U$ becomes comparable and these two competing interactions give rise to emergent phenomena. 

For the experimenters and theoreticians alike, the family of iridates, one of the $5d$ oxide families, have become a test bed for exploring these emergent phenomena that includes topological Mott insulator \cite{Pesin2010}, fractional Chern insulator \cite{Bergholtz2013}, Kitaev's celebrated spin liquid phase \cite{Baskaran2007,Takagi2019,Okamoto2007}, Dirac and Weyl semimetals \cite{Yan2017,Vafek2014}, Axion insulator \cite{Wan2012}, and high-$T_\mathrm{c}$ superconductivity. For example, pyrochlore iridates exhibit topological semimetal and Mott insulating phases \cite{Wan2011,Krempa2013,Topp2018}. The honeycomb iridates host unconventional magnetic and spin liquid phases \cite{Rau2014,Chaloupka2013,Reuther2014,Takasashi2019,Kenney2019}. The monolayer and bilayer Ruddlesden-Popper strontium iridates, Sr$_2$IrO$_4$ and Sr$_3$Ir$_2$O$_7$, exhibits canted in-plane and collinear out-of-plane antiferromagnetic (AF) insulating state \cite{Nauman2017,Kim2008,Wantabe2013,Yan2015,Fujiyama2012}. 

Among the family of iridates, the orthorhomic pervoskite iridates with chemical formula AIrO$_3$, constitutes a large class of anisotropic oxides with exceptionally wide range of properties. In these oxides the IrO$_6$ octahedra is distorted as compared to the symmetric cubic structure. These class of oxides with Pbnm space group symmetry was proposed to realize a new class of metal dubbed topologically crystalline metal with flat surface state and phase transition to other non-trivial phases in the presence of external field has been theoretically proposed \cite{Chen2015}. For example, SrIrO$_3$ and CaIrO$_3$ are paramagntic Dirac semimetals (DSM) \cite{Nie2015,Fujioka2017,Fujioka2019,Fujioka2021}. Carter \textit{et al.} \cite{Carter2012} showed that SrIrO$_3$ undergoes a phase transition from semimetal to strong topological insulator by breaking the mirror symmetry. BaIrO$_3$ is a Pauli paramagnetic metal (PM) with Fermi liquid ground state below 6 K \cite{Cheng2013}. In comparison to the AIrO$_3$, the other $3d$ and $4d$ perovskites with isolvalent $d^5$ configuration, where the SOC is weak and electron correlation is strong, have different electronic and magnetic phases in their ground state. For example, SrCoO$_3$ and CaCoO$_3$ exhibits high-spin (HS) ferromagnetic metal (FM) state whereas SrRhO$_3$ and YRuO$_3$ exhibits low-spin (LS) FM and canted antiferromagnetic insulator (CAFI) ground state \cite{Xia2017,MathiJaya1991,Singh2003,Ji2020}.

For the perovskites, due to strong octahedral crystal field the $d$ orbital degeneracy gets lifted giving rise to $t_\mathrm{2g}$ and $e_\mathrm{g}$ manifold. In the presence of SOC the $t_\mathrm{2g}$ manifold splits into spin-orbital entangled pseudo-spin $J_\mathrm{eff}$ = 3/2 and $J_\mathrm{eff}$ = 1/2 states. In the case of a $d^5$ (Ir$^{4+}$) configuration, the $J_\mathrm{eff}$ = 3/2 states are completely occupied and the ground state is constituted of the $J_\mathrm{eff}$ = 1/2 pseudo-spin. These narrow $J_\mathrm{eff}$ = 1/2 bands formed due to strong SOC are more susceptible to Mott localization by $U$, due to reduced bandwidth, which implies that stronger the SOC is, weak critical $U$ ($U_\mathrm{c}$) will be required to make a transition from metal to a spin-orbit coupled Mott insulator. By carrying out
first-principles density functional theory (DFT) calculations within the framework of local density approximation (LDA)+$U$ on the prototype system SrIrO$_3$, Zeb $\textit{et al.}$ \cite{Zeb2012} presented an electronic phase diagram in the $U$-SOC configuration space, and broadly suggested three domains. These are magnetic metal in the weak to intermediate SOC and weak to high $U$ regime, nonmagnetic metal (NM) or DSM in the weak to intermediate $U$ and weak to high SOC regime, and magnetic insulator in the weak to high SOC and intermediate to high $U$ regime. \par 

The aforementioned discussion on the electronic structure of orthorhombic perovskites suggest that a much more intricate phase diagram needs to be constructed for the family of $d^5$ oxides in the $U$-SOC space in order to explore the emerging quantum phases in this family. Firstly, it needs to be revisited if in the absence of SOC, the system remains metallic even for very high $U$. Secondly, in the absence of collinear spin-arrangement questions arise whether sub-domains with non-trivial quantum states exist within the four proposed broad domains. Thirdly, the role of structural distortions - from the isotropic cubic configuration to symmetry lowered orthorhombic configuration -  needs to be examined in building the phase diagram as it is known to be a key player in driving the electronic and magnetic phase transitions in the perovskite family \cite{Torrance1992,Garc1992,Miyasaka2003,Zhou2005,Zhou2010,Vogt2003,Kim2017,Liu2015}.

In this work we have employed a multi-band Hubbard model which is solved self-consistently in the mean-field framework and validated the results with DFT+$U$ calculations to establish a detailed phase diagram discovering the earlier unexplored sub-domains. The present study establishes the complex interplay among SOC, $U$, $\Delta$ and structural distortions to explain the formation of collinear and non-collinear magnetism as well as metallic, insulating, and non-trivial semiemtallic phases in the family of LS $d^5$ perovskites. 

\section{Structural and Computational Details}
\begin{figure}
\centering
\includegraphics[angle=-0.0,origin=c,height=8cm,width=9cm]{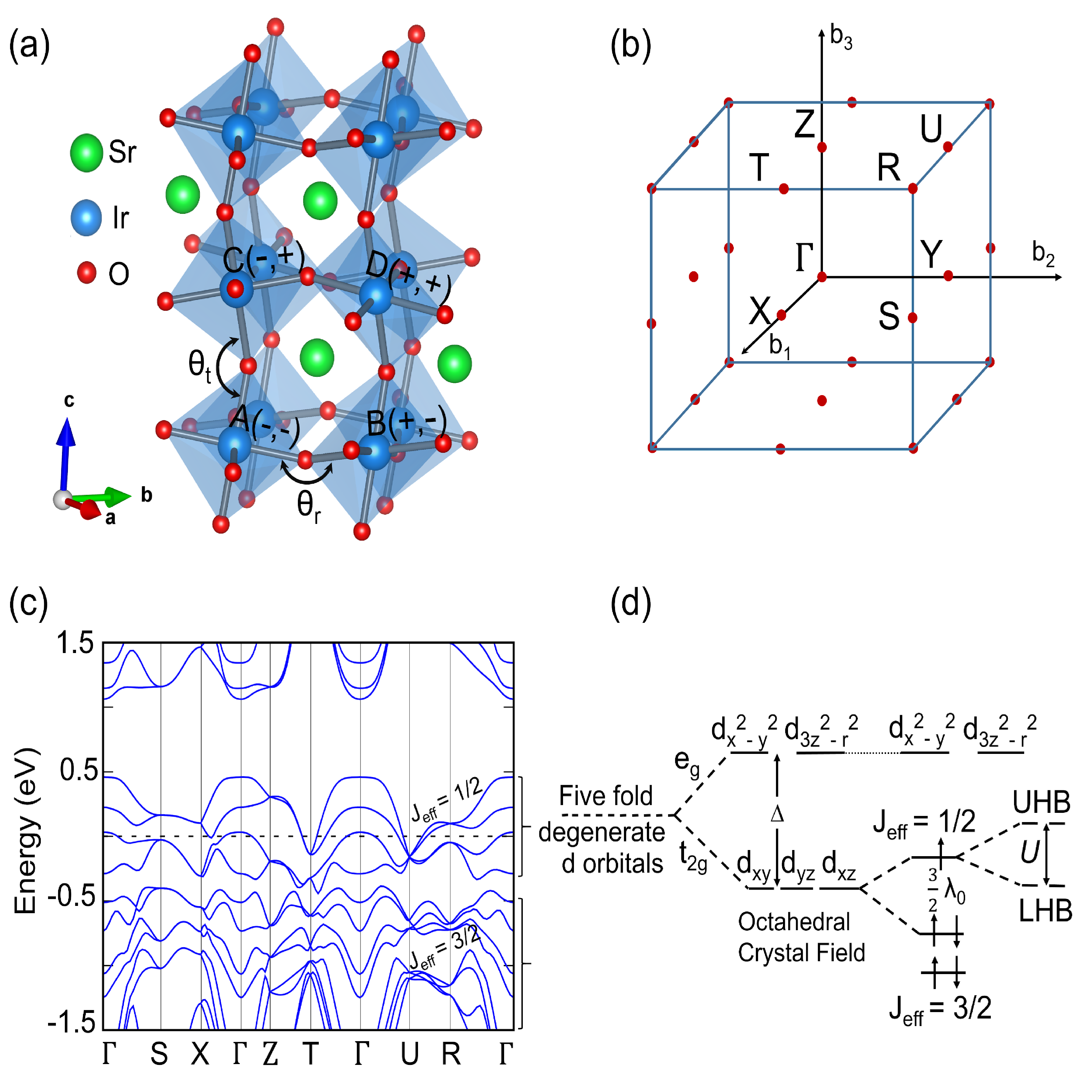}
\caption{(a) Crystal structure of orthorhombic SrIrO$_3$.(b) Corresponding bulk Brillouin zone with high symmetric k-points. (c) Bulk electronic structure of SrIrO$_3$ obtained using DFT+$U$ with $U_\mathrm{eff}$ = 0 and $\lambda$/$\lambda_0$ = 1. Here, $\lambda_0$ is the real SOC strength with magnitude 0.43 eV, respectively. The Fermi level is set to be 0. A Dirac node appears at the high symmetric k-point U revealing the semimetallic nature of SrIrO$_3$. (d) Schematic illustration of the effect of crystal field, SOC and onsite Coulomb repulsion on the Ir-$d$ states.}
\label{Fig1}
\end{figure}

Bulk SrIrO$_3$ crystallizes in an orthorhombic pervoskite crystal structure (space group Pbnm) with the GdFeO$_3$-type lattice distortion as shown in Fig. \ref{Fig1}(a). The IrO$_6$ octahedra undergoes a staggered rotation about the $c$ axis by an angle $\theta_\mathrm{r}$ =  ${153.53}^{\circ}$, followed by another rotation about the [110] direction by tilt angle $\theta_\mathrm{t}$ = ${156.52}^{\circ}$. Due to these tilt and rotations of the IrO$_6$ octahedra, the unit cell gets doubled in the $\textit{ab}$ as well as in the $\textit{ac}$ plane due to $\sqrt{2}a_\mathrm{0} \times \sqrt{2}a_\mathrm{0} \times 2a_\mathrm{0}$ supercell geometry where $a_\mathrm{0}$ is the nearest-neighbour Ir-Ir distance. Hence, SrIrO$_3$ contains four formula units per unit cell leading to four inequivalent Ir sublattices (A,B,C,D). The two signs on each Ir sublattice represents the sense of the rotation and tilting of the octahedra, clockwise (+) or anticlockwise (-). The experimental lattice parameters are $a$ = 5.56 {\AA}, $b$ = 5.59 {\AA} and $c$ = 7.88 {\AA} \cite{Zhao2008}.
In order to study the effect of correlation and SOC on the electronic and magnetic phases in bulk SrIrO$_3$, we developed a multi-band Hubbard model and solved it self-consistently as discussed in Sec. \ref{IV} and also carried out comprehensive DFT calculations in the $U$-SOC space. The calculations are performed using plane-wave based projector augmented wave method (PAW) \cite{Kresse1999,Blochl1994} as implemented in Vienna ab-initio simulation package (VASP) \cite{Kresse1996} within the Perdew$-$Burke$-$Ernzerhof generalized gradient approximation (GGA) for exchange-correlation functional. In this study, the SOC $\lambda$ is varied in units of $\lambda_\mathrm{0}$, where $\lambda_\mathrm{0}$ is the real SOC strength (= 0.43 eV) as obtained from the SCF calculations to achieve the ground state. 
The Brillouin zone integrations are carried out using $8 \times 8 \times 4$ Monkhorost-Pack k-mesh which yields 256 k-points in the irreducible part of the Brilluion zone. The kinetic energy cutoff for plane-wave basis set was chosen to be $500$ eV. The strong correlation effect is incorporated via an effective onsite correlation parameter $U_\mathrm{eff}$ = $U-J$ through the rotationally invariant approach introduced by Dudarev \cite{Dudarev1998}.

\section{Ground State Electronic Structure of SrIrO$_3$}
The ground state electronic structure, as shown in Fig. \ref{Fig1}(c), implies that bulk SrIrO$_3$ exhibits nonmagnetic semimetallic ground state with a Dirac node at the high symmetry point U. This is due to the combined effect of cubic asymmetry, octahedral crystal field and SOC towards the removal of degeneracy in the $d$ manifold as illustrated in Fig. \ref{Fig1}(d). Due to strong octahedral crystal field, the five-fold degenerate $d$ states split into a higher-energy and unoccupied $e_\mathrm{g}$ doublet and a lower-energy $t_\mathrm{2g}$ triplet. While the $e_\mathrm{g}$ states are unaffected by the SOC, the $t_\mathrm{2g}$ manifold is further SOC split into a  $J_\mathrm{eff}$ = 1/2 doublet ($m_\mathrm{j}$ = $\pm$ {1/2}) and a  $J_\mathrm{eff}$ = 3/2 ($m_\mathrm{j}$ = $\pm$ {3/2}, $\pm$ {1/2}) quartet, hence, forming three Kramer's pair. The expressions $\ket{J,m_\mathrm{j}}$ are given by,

\begin{align}
  \ket{\frac{1}{2}, \pm{\frac{1}{2}}} &= \frac{1}{\sqrt{3}}( \ket{yz,\bar{\sigma}} \pm{\ket{xy,\sigma}} \pm{i} \ket{xz,\bar{\sigma}})\\
   \ket{\frac{3}{2}, \pm{\frac{1}{2}}} &= \frac{1}{\sqrt{6}}(\ket{yz,\bar{\sigma}} \mp 2{\ket{xy,\sigma} } \pm{i} \ket{xz,\bar{\sigma}})\\
    \ket{\frac{3}{2}, \pm{\frac{3}{2}}} &= \frac{1}{\sqrt{2}}(\ket{yz,\sigma} \pm{i} \ket{xz,{\sigma}})
  \end{align}
  
where $\pm$ corresponds to spin $\sigma$ =  $\uparrow$/$\downarrow$, respectively. With lowering in symmetry through $\theta_\mathrm{r}$ and $\theta_\mathrm{t}$, two pairs of bands make linear crossing at the high symmetry point U to create double Dirac nodes resulting in a DSM phase. It has been reported that the two Dirac nodes may not be degenerate in energy and in that case the cones may penetrate to form a nodal ring \cite{Zeb2012}. Through our model Hamiltonian in the Sec. \ref{IV}, we will attribute the formation of the DSM phase due to the development of $t_\mathrm{2g}$-$e_\mathrm{g}$ interaction in the next-nearest neighborhood due to tilting and rotation of octahedra. Besides the crystal field and SOC, the onsite Coulomb repulsion is also deterministic of the electronic structure of the $d^5$ (Ir$^{4+}$) compounds. Four out of five $d$ electrons occupy the  $J_\mathrm{eff}$ = 3/2 state, leaving one electron in the  $J_\mathrm{eff}=1/2$ state (see Fig. \ref{Fig1}(d)). The onsite repulsion further splits this degenerate state to create a lower Hubbard (LHB) and upper Hubbard (UHB) subband. However, experimental studies \cite{LONGO1971,Nie2015,Fujioka2017} including ARPES measurements show that the SrIrO$_3$ is a paramagnetic semimetal. This demonstrates that the competition between SOC and $U$ can lead to a plethora of electronic and magnetic phases for the $d^5$ perovskites.

Apart from the perovskite structure, several other iridates and iridate like systems with $d^5$ configuration and with similar octahedral complexes, exhibit wide range of magnetic phases with spin anisotropy \cite{Modic2014,Nauman2017,Boeggia2013,Chaloupa2016,Biffin2014,Reuther2014}. Therefore, we utilize SrIrO$_3$ as prototype to build a platform so that the emerging quantum phases in the $U$-SOC configuration space can be envisaged and analyzed. The phase diagram will be discussed in detail in the Sec. \ref{VA}.

\section{Multi-Band Hubbard Model}
\label{IV}
Across the iridates family, it has been shown that the competition between $U$ and $\lambda$ influences the eigen states substantially. To get a flavour of it in the single pervoskite structure, we employ a mean-field based multi-band Hubbard model Hamiltonian to explore the non-trivial/trivial phases emerging in the weak/strong $U$-SOC regime. To start with, we first consider tight binding (TB) model with SOC, where a minimal basis set formed by Ir-$d$ orbitals ($xy$,$yz$,$xz$,$x^2-y^2$,$3z^2-r^2$) has been considered. The five orbital basis set, instead of $t_\mathrm{2g}$ based three orbital, introduces $t_\mathrm{2g}$-$e_\mathrm{g}$ intermixing due to finite $\theta_\mathrm{r}$ and $\theta_\mathrm{t}$ in the distorted frame leading to significant altering of the band structure. 

To extract the nearest-neighbour (NN), next-nearest-neighbour (NNN) $\sigma$ and $\pi$ hopping interactions and SOC strength, a TB model for undistorted SrIrO$_3$ is formulated and then using these parameters and transforming the hopping matrices in the rotated basis (as explained in Appendix \ref{A}), the effect of distortion on the electronic and magnetic properties has been examined. 
In the second quantization notation, the TB + SOC component of the Hamiltonian is given by,
\begin{eqnarray}
    H_\mathrm{TB-SOC}&=&\sum_{i,\alpha}\epsilon_{i,\alpha} c_{i,\alpha}^\dagger c_{i,\alpha} + \nonumber \\  && \sum_{i,j,\alpha,\beta,\sigma}t_{i,\alpha,j, \beta}(c_{i,\sigma,\alpha}^\dagger c_{j,\sigma,\beta} + h.c.) \nonumber \\
    &&+ \lambda \sum_{\alpha,\beta,\sigma,\bar{\sigma}} \bra{\alpha \sigma} \boldsymbol{L} \cdot \boldsymbol{S} \ket{\beta \bar{\sigma}} c_{\alpha, \sigma}^\dagger c_{\beta,\bar{\sigma}}
\end{eqnarray}

Here, i(j), $\alpha$($\beta$) are site and orbital indices, respectively. The parameters $\epsilon_\mathrm{i\alpha}$ and $t_\mathrm{i \alpha j \beta}$ represents the onsite energy and strength of hopping integrals, respectively. The SOC is added in the third term of the Hamiltonian with $\lambda$ denoting its strength. In compact form, for a single formula unit as is the case with cubic perovskite structure, H is given by

\begin{equation}
    H= \begin{pmatrix}
       H_{\uparrow \uparrow}^{5\times5} & H_{\uparrow \downarrow }^{5\times5}\\[0.2cm]
       H_{\downarrow \uparrow}^{5\times5} & H_{\downarrow \downarrow}^{5\times5}
       \end{pmatrix} 
\end{equation}

Here, H$_{\uparrow \uparrow}$ = H$_{\downarrow \downarrow}$ and H$_{\uparrow \downarrow}$ = (H$_{\downarrow \uparrow})^\dagger$ to ensure the time reversal (TR) invariance of the Hamiltonian. For the four formula unit ($\sqrt{2}a_\mathrm{0} \times \sqrt{2}a_\mathrm{0} \times 2a_\mathrm{0}$) supercell, which build the primitive unit cell of the orthorhombioc phase, the augmented Hamiltonian representing four-Ir (A,B,C,D) sublattices then takes the shape of
\begin{equation}
    H= \begin{pmatrix}
       H_{\uparrow \uparrow}^{20\times20} & H_{\uparrow \downarrow}^{20\times20}\\[0.2cm]
       H_{\downarrow \uparrow}^{20\times20} & H_{\downarrow \downarrow}^{20\times20}
       \end{pmatrix} 
\end{equation}

The TB component built with Slater-Koster formalism \cite{Slater1954} is further elaborated in the Appendix \ref{A}.

For the distorted case, with finite $\theta_\mathrm{r}$ and $\theta_\mathrm{t}$, as defined through Fig. \ref{Fig1}(a), the hopping matrices are obtained by using the following transformation 

\begin{align}
    \tilde{H}_\mathrm{TB}&= R^T H_\mathrm{TB} R
\end{align}

where R is the transformation matrix and can be calculated using rotation matrix for the cubic harmonics (L = 2) (Eq. A3 of Appendix \ref{A}). For the rotation and tilting, as appropriate for SrIrO$_3$, it is respectively defined as

\begin{figure}[hbt] 
\centering
\includegraphics[angle=-0.0,origin=c,height=7cm,width=8.5cm]{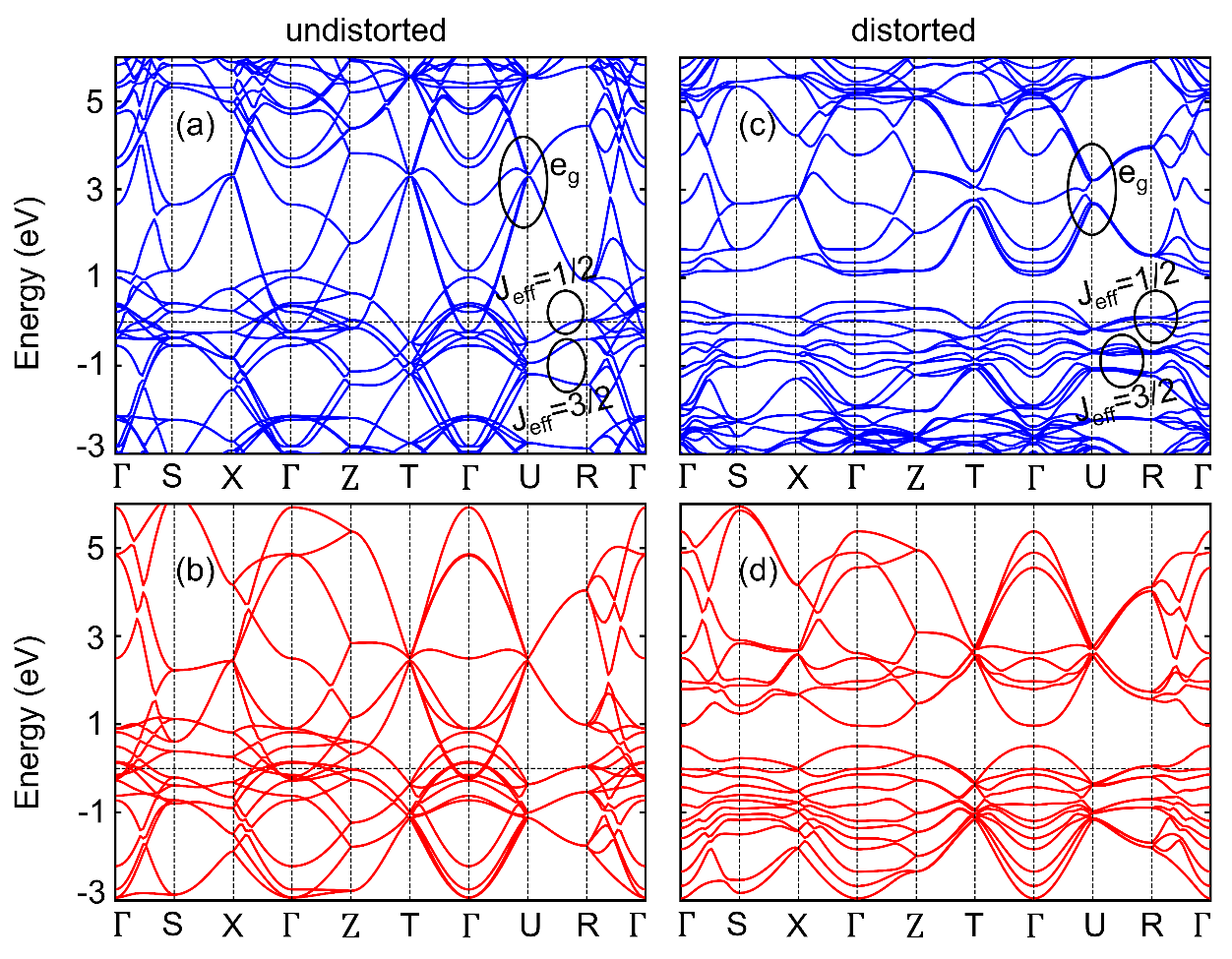}
\caption{DFT (blue) and TB (red) band structures for undistorted (a,b) and distorted (c,d) SrIrO$_3$ with $\lambda$ = 0.43 eV, respectively. Segregation of $t_\mathrm{2g}$ and $e_\mathrm{g}$ states, as well as the formation of DSM phase, due to distortion of the octahedra.}
\label{TB_DFT_bands}
\end{figure}

\begin{widetext}
 \begin{equation}
       R(\theta_\mathrm{t}) = \begin{pmatrix}
       \frac{1}{2}(1+\cos^2{\theta_\mathrm{t}}) & \frac{1}{2\sqrt{2}}\sin{2\theta_\mathrm{t}} & -\frac{1}{2\sqrt{2}}\sin{2\theta_\mathrm{t}} & 0 & -\sqrt{\frac{3}{4}}\sin^2{\theta_\mathrm{t}}  \\[0.2cm]
       
       -\frac{1}{2\sqrt{2}}\sin{2\theta_\mathrm{t}} & \frac{1}{2}(2\cos^2\theta_\mathrm{t}+\cos{\theta_\mathrm{t}}-1) & -\frac{1}{2}(2\cos^2{\theta_\mathrm{t}} -\cos{\theta_\mathrm{t}}-1) & -\frac{1}{\sqrt{2}}\sin{\theta_\mathrm{t}} & -\sqrt{\frac{3}{8}}\sin{2\theta_\mathrm{t}} \\[0.2cm]
       
      \frac{1}{2\sqrt{2}}\sin{2\theta_\mathrm{t}} & -\frac{1}{2}(2\cos^2{\theta_\mathrm{t}} -\cos{\theta_\mathrm{t}}-1) & \frac{1}{2}(2\cos^2{\theta_\mathrm{t}} +\cos{\theta_\mathrm{t}}-1) & -\frac{1}{\sqrt{2}}\sin{\theta_\mathrm{t}}  & \sqrt{\frac{3}{8}}\sin{2\theta_\mathrm{t}}  \\[0.2cm]
       
       0 & \frac{1}{\sqrt{2}}\sin{\theta_\mathrm{t}} & \frac{1}{\sqrt{2}}\sin{\theta_\mathrm{t}} & \cos{\theta_\mathrm{t}} & 0\\[0.2cm]
       
       -\sqrt{\frac{3}{4}}\sin^2{\theta_\mathrm{t}} & \sqrt{\frac{3}{8}}\sin{2\theta_\mathrm{t}} & -\sqrt{\frac{3}{8}}\sin{2\theta_\mathrm{t}} & 0 & 1-\frac{3}{2}\sin^2{\theta_\mathrm{t}} 
       \end{pmatrix} 
\end{equation}
\end{widetext}

\begin{equation}
    R(\theta_\mathrm{r}) = \begin{pmatrix}
       \cos{2\theta_\mathrm{r}} & 0 & 0 & \sin{2\theta_\mathrm{r}} & 0  \\[0.2cm]
       0 & \cos{\theta_\mathrm{r}} &  \sin{\theta_\mathrm{r}} & 0 & 0 \\[0.2cm]
       0 & -\sin{\theta_\mathrm{r}} &  \cos{\theta_\mathrm{r}} & 0 & 0 \\[0.2cm]
      -\sin{2\theta_\mathrm{r}} & 0 & 0 & \cos{2\theta_\mathrm{r}} & 0  \\[0.2cm]
       0 & 0 & 0 & 0 & 1  \\[0.2cm]
       \end{pmatrix} 
\end{equation}
The values of onsite energy ($\epsilon$), hopping strength ($t$) and SOC strength ($\lambda_\mathrm{0}$) are obtained by fitting the TB+SOC bands with the DFT+SOC bands for the undistorted case as shown in Figs. \ref{TB_DFT_bands} (a,b). The fitted values are listed in Table \ref{tb par}. TB bands, as shown in Fig. \ref{TB_DFT_bands}, captures very well the essential features of the DFT band structure. In this LS ($t_\mathrm{2g}^5e_\mathrm{g}^0$) configuration, while the overlapping of the $t_\mathrm{2g}$ and $e_\mathrm{g}$ bands is prominent in the undistorted frame (Figs. \ref{TB_DFT_bands}(a,b)), they are very well segregated in the distorted frame (Fig. \ref{TB_DFT_bands}(c)) and this is very well captured in our TB model (Fig. \ref{TB_DFT_bands}(d)). The distortion also introduces a DSM phase as linear bands cross each other at the high symmetry point U in the vicinity of the Fermi energy ($E_\mathrm{F}$).

\begin{table}[hbt]
\caption{Calculated values of onsite energy, interaction parameters and SOC strength in units of eV. The parameters $\epsilon_\mathrm1$ and $\epsilon_\mathrm{2}$ give the values of onsite energy for $t_\mathrm{2g}$ and $e_\mathrm{g}$ states, respectively. For the undistorted case, we have considered identical hopping strengths, $t_\mathrm{1}$ (NN $\pi$), $t_\mathrm{2}$ (NNN $\sigma$) and $t_\mathrm{3}$ (NNN $\pi$) for $t_\mathrm{2g}$ - $t_\mathrm{2g}$ interactions, $t_\mathrm{4}$ (NN $\sigma$), $t_\mathrm{5}$ (NNN $\sigma$), $t_\mathrm{6}$ (NNN $\pi$) for $e_\mathrm{g}$-$e_\mathrm{g}$ interactions and $t_\mathrm{7}$ (NNN $\sigma$) for $t_\mathrm{2g}$-$e_\mathrm{g}$ interactions, as relevant for the cubic symmetry.}
\vspace{0.2cm}
\begin{tabular}{c c c c c c c c c c}
\hline \hline
$\epsilon_\mathrm{1}$ & $\epsilon_\mathrm{2}$ & $t_\mathrm{1}$ & $t_\mathrm{2}$ & $t_\mathrm{3}$ & $t_\mathrm{4}$ & $t_\mathrm{5}$ & $t_\mathrm{6}$ & $t_\mathrm{7}$ & $\lambda$  \\
\hline
-0.79 & 2.4 & -0.38 & -0.13  & 0.04 & -0.85 & -0.5 & 0.1 & 0.04 & 0.43\\
\hline \hline
\end{tabular}
\label{tb par}
\end{table}

Having formulated the kinetic part of the Hamiltonian, we now consider the interacting part of the Hamiltonian ($H_\mathrm{int}$) which is described in terms of the multi-orbital Hubbard-Kanamori formalism \cite{Luo2013}.  
\begin{equation}
\begin{split}
    H_\mathrm{int} = U \sum_{i,\alpha} n_\mathrm{i,\alpha,\uparrow} n_\mathrm{i,\alpha,\downarrow}
    + (U^\prime -\frac{J_\mathrm{H}}{2}) \sum_{i,\alpha < \beta} n_\mathrm{i,\alpha} n_\mathrm{i,\beta}\\
    -2J_\mathrm{H} \sum_{i,\alpha < \beta} S^z_\mathrm{i,\alpha} \cdot S^z_\mathrm{i,\beta} = H_1 + H_2 + H_3
\end{split}
\end{equation}

Here, the first two terms gives the energy cost of having the electrons in the same or different orbitals at the same lattice site. The third term defines the Hund's rule coupling that favours the ferromagnetic alignment of spins in the orbitals at the same lattice site. The relation $U^\prime$ = $U$ - $2J_\mathrm{H}$ between the Kanamori parameters \cite{Kanamori1963} has been used here.

In the Hartree approximation, $H_\mathrm{1}$, $H_\mathrm{2}$, and $H_\mathrm{3}$ can be decoupled as

\begin{align}
H_{1}\approx U \sum_{i,\alpha} (n_\mathrm{i,\alpha,\uparrow}\langle  n_\mathrm{i,\alpha,\downarrow}\rangle  
+n_\mathrm{i,\alpha,\downarrow}\langle  n_\mathrm{i,\alpha,\uparrow}\rangle \notag\\
-\langle  n_\mathrm{i,\alpha,\downarrow}\rangle \langle n_\mathrm{i,\alpha,\uparrow}\rangle) 
\end{align}

\begin{align}
H_{2} \approx U \sum_{i,\alpha < \beta} (n_\mathrm{i,\alpha}\langle  n_\mathrm{i,\beta}\rangle  
+n_\mathrm{i,\beta}\langle  n_\mathrm{i,\alpha}\rangle 
-\langle  n_\mathrm{i,\alpha}\rangle  \langle  n_\mathrm{i,\beta}\rangle  )
\end{align}

\begin{equation}
\begin{split}
H_{3} \approx -\frac{J_H}{2}\sum_{i,\alpha < \beta} ( n_\mathrm{i,\alpha,\uparrow}\langle n_\mathrm{i,\beta,\uparrow}\rangle  
+ n_\mathrm{i,\beta,\uparrow}\langle n_\mathrm{i,\alpha,\uparrow}\rangle \\
- n_\mathrm{i,\alpha,\uparrow}\langle n_\mathrm{i,\beta,\downarrow}\rangle  
- n_\mathrm{i,\beta,\downarrow}\langle n_\mathrm{i,\alpha,\uparrow}\rangle\\
- n_\mathrm{i,\alpha,\downarrow}\langle n_\mathrm{i,\beta,\uparrow}\rangle  
- n_\mathrm{i,\beta,\uparrow}\langle n_\mathrm{i,\alpha,\downarrow}\rangle \\
+ n_\mathrm{i,\alpha,\downarrow}\langle n_\mathrm{i,\beta,\downarrow}\rangle
+ n_\mathrm{i,\beta,\downarrow}\langle n_\mathrm{i,\alpha,\downarrow}\rangle \\
- \langle n_\mathrm{i,\alpha,\uparrow}\rangle \langle n_\mathrm{i,\beta,\uparrow}\rangle 
+ \langle n_\mathrm{i,\alpha,\uparrow}\rangle \langle n_\mathrm{i,\beta,\downarrow}\rangle \\
+ \langle n_\mathrm{i,\alpha,\downarrow}\rangle \langle n_\mathrm{i,\beta,\uparrow}\rangle  
- \langle n_\mathrm{i,\beta,\downarrow}\rangle\langle n_\mathrm{i,\alpha,\downarrow}\rangle )
\end{split}
\end{equation}

where, $n_\mathrm{i,\alpha}$ = $n_\mathrm{i,\alpha,\uparrow}$ + $n_\mathrm{i,\alpha,\downarrow}$, is the total charge density of orbital $\alpha$ at site i.

Hence, the total Hamiltonian is given by 
\begin{align}
H = H_\mathrm{TB-SOC} + H_\mathrm{int}.
\end{align}

\begin{figure}[H]
\centering
\includegraphics[angle=-0.0,origin=c,height=11cm,width=9cm]{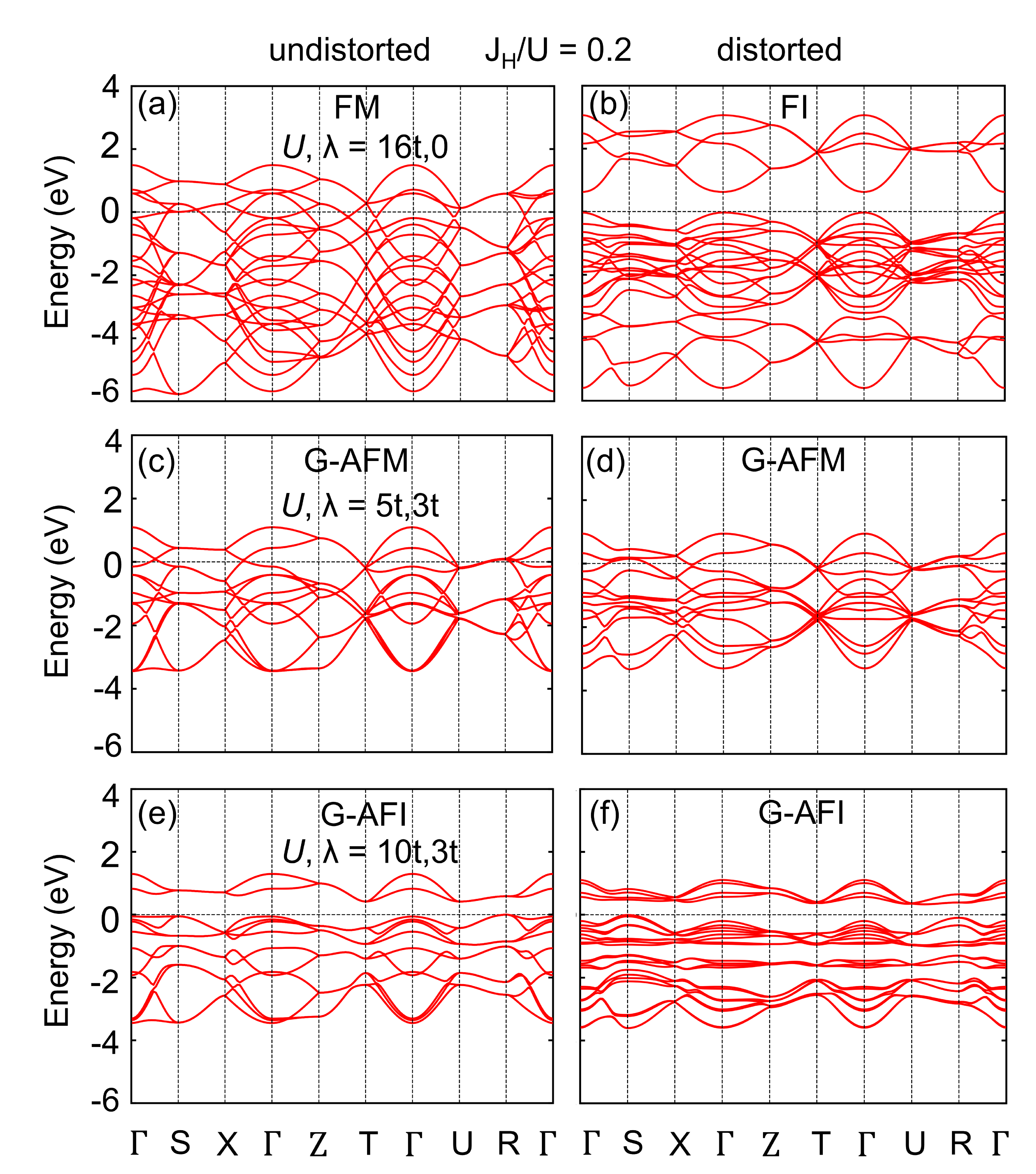}
\caption{Electronic band structures obtained for the undistorted structure (first column) and for the distorted structure (second coloumn) with $t$ = 0.38 eV defined as the energy scale unit. (a,b) without SOC and high $U$, (c,d) with SOC and low $U$, and (e,f) with SOC and moderate $U$ value.}
\label{mean_field_bands}
\end{figure}

It is solved  self-consistently in the momentum space by employing Hartree approximation as described above. The $U$-SOC space give rise to different magnetic phases, as shown in Fig. \ref{mean_field_bands}, beyond the semimetallic nonmagnetic phase. We set $J_\mathrm{H}$/$U$ = 0.2 as relevant for $4d/5d$ oxides \cite{Meetei2015} and $t$ = $t_\mathrm{1}$ as the energy scale unit. For $\lambda$ = 0, a FM phase (see Fig. \ref{mean_field_bands}(a)), with highly dispersive bands (higher mobility) stabilizes. For even higher $U$ values, a FM state remains stable for the undistorted case.

With SOC, the system shows intriguing magnetic phases where the $J_\mathrm{eff}$ = 1/2 states determine the ground state. For lower $U$ values, a weak G-type antiferromagnetic metal (G-AFM) state (Fig. \ref{mean_field_bands}(c)) with broken Dirac node forms the ground state. It is due to breaking of the TR symmetry. Here, the SOC induces the spin degeneracy and strengthen the localization and hence decreases the mobility. The weak G-AFM state has small hole and electron pockets which disappears on further increasing $U$ and a gap is opened in the $J_\mathrm{eff}$ = 1/2 spectrum to stabilize a G-type antiferromagnetic insulator (G-AFI) phase (Fig. \ref{mean_field_bands}(e)). Finite distortion induces anisotropy in the orbital ($t_\mathrm{2g}$) occupancies with onsite Coulomb repulsion enhances the spin split and drives the system to a ferromagnetic insulator (FI) phase (Fig. \ref{mean_field_bands}(b)). Hence, competition between $U$, SOC and structural distortion alters the magnetic and electronic phase of the system substantially. Recent studies on iridates and their superlattices emphasize the crucial role of distortion induced anisotropic spin interactions leading to non-collinear magnetic phases \cite{Mohapatra2019,Kim2017,Liu2015}. However, the current model is based on collinear magnetism and therefore cannot predict non-collinear phases due to the absence of exchange terms in the Hamiltonian. Therefore, we carried out pseudopotential based DFT calculations to explore the novel electronic and magnetic states in the $U$-SOC domain and build the quantum phase diagram of $d^5$ pervoskites in the LS state by taking SrIrO$_3$ as a prototype.

\section{DFT+$U$+SOC Electronic Structure}
\begin{center}
\begin{figure}[H]
\centering
\includegraphics[angle=-0.0,origin=c,height=8cm,width=18cm]{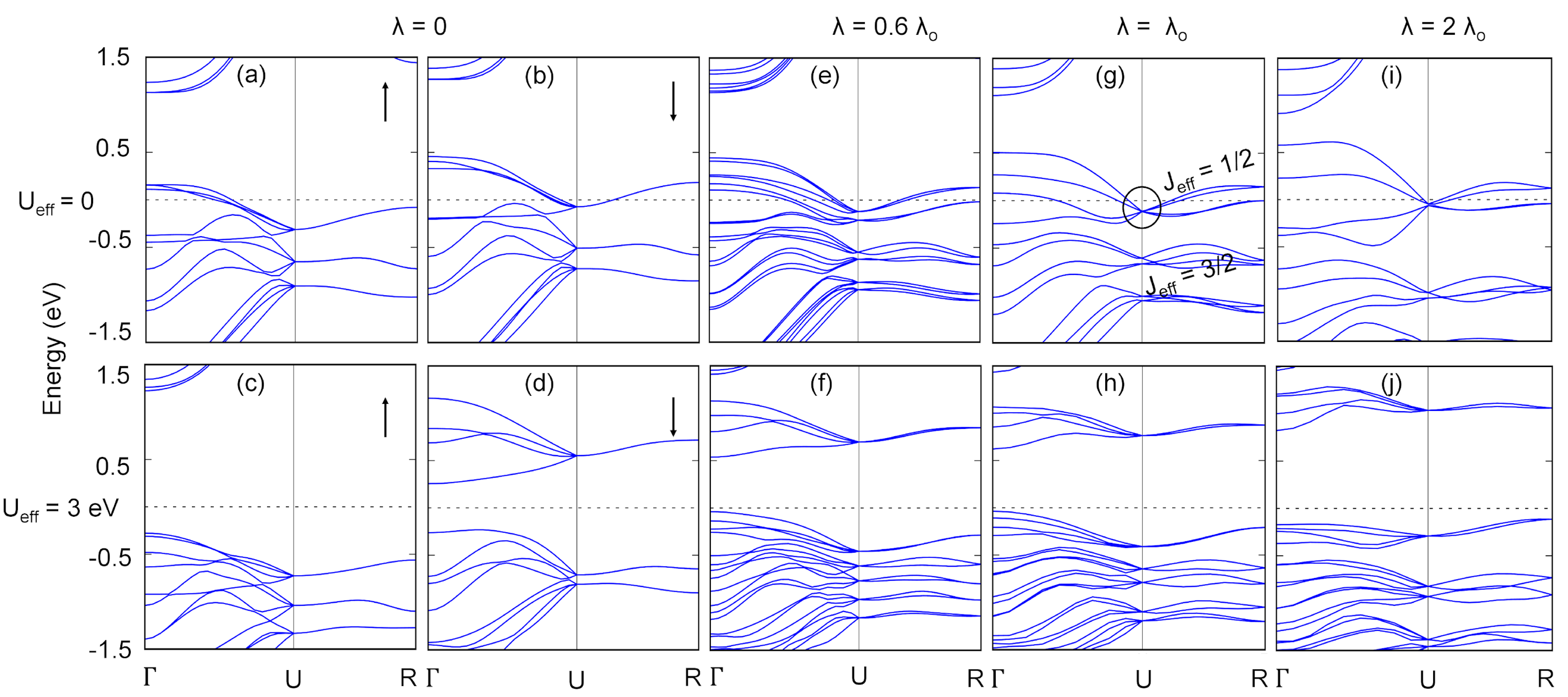}
\caption{Evolution of bulk electronic structure as a function of $U_{eff}$ and SOC. The upper and lower rows represents the band structure without and with $U_{eff}$ for four values of $\lambda$/$\lambda_0$, namely, 0 (a - d), 0.6 (e, f), 1 (g, h), and 2 (i, j). (Upper row) Mixing of up-spin and down-spin states with increasing $\lambda$/$\lambda_0$ leading to the formation of DSM phase. (Lower row) Amplified effect of correlations with increasing  $\lambda$/$\lambda_0$. Enhanced spin splitting leads to a gap opening and stabilize the system in a ferromagnetic insulator state for $\lambda$ = 0 and in a canted antiferromagnetic insulator state for finite $\lambda$/$\lambda_0$.}
\label{Fig4}
\end{figure}
\end{center}

Through Fig. \ref{Fig4}, we will analyze the electronic structure evolution as a function of $U$ and SOC. In the ground state, we estimated the SOC strength $\lambda_\mathrm{0}$ to be 0.43 eV by measuring the split between $J_\mathrm{eff}$ = 1/2 and $J_\mathrm{eff}$ = 3/2 states which is equal to 3/2 $\lambda_\mathrm{0}$ (see Fig. \ref{Fig1}(d)). In order to examine the role of $\lambda$, we scaled it in units of $\lambda_0$. For $\lambda/\lambda_0$ = $U_{eff}$ = 0, it is pure octahedral crystal field effect where the $d$ orbitals split into lower-lying and partially occupied three-fold degenerate $t_\mathrm{2g}$ and upper-lying two-fold degenerate $e_\mathrm{g}$ states as shown in Fig. \ref{Fig4}(a,b). Due to imbalance in the population of the states in two different spin channels the system becomes a Stoner ferromagnet. As $\lambda/\lambda_0$ increases, real spin states of the $t_\mathrm{2g}$ manifold evolve and give rise to pseudo-spin states with the formation of spin-orbit entangled $J_\mathrm{eff}$ = 1/2 states lying in the vicinity of the $E_\mathrm{F}$ and $J_\mathrm{eff}$ = 3/2 states lying below in the valence band. For weak $\lambda/\lambda_0$, the four $J_\mathrm{eff}$ = 1/2 pairs, corresponding to the four-Ir sublattices, create two set of four-fold degenerate bands along the k-path U-R. These two sets merge at the high symmetry point U with increasing SOC to form a DSM phase (see Figs. \ref{Fig4}(e,g,i)).

With increasing correlation effect, say for $U_{eff}$ = 3 eV and no SOC, as shown in Figs. \ref{Fig4}(c,d), the onsite repulsion $U$n$_{\uparrow}$n$_{\downarrow}$ increases the spin split to form a gap which in turn makes the system a FI. This phase is very well captured by the model Hamiltonian described in the previous section. As SOC competes with $U$, Figs. \ref{Fig4}(f,h,j), the four $J_\mathrm{eff}$ = 1/2 pairs create lower and upper Hubbard subbands, which breaks the Dirac node to form a gap and concurrently stabilizes the system in a CAFI state. The band gap rises from 0.23 eV at $\lambda$/$\lambda_\mathrm{0} = 0$  to 0.93 eV at $\lambda$/$\lambda_\mathrm{0}$ = 2, manifesting the amplified effect of correlations as the SOC strength increases. The spin anisotropy and the overall phases predicted in the $U$-SOC domain will be elaborated further in the next subsection.

\onecolumngrid
\twocolumngrid

\subsection* {BULK PHASE DIAGRAM}
\label{VA}
\begin{figure}
\centering
\includegraphics[angle=-0.0,origin=c,height=13.8cm,width=9cm]{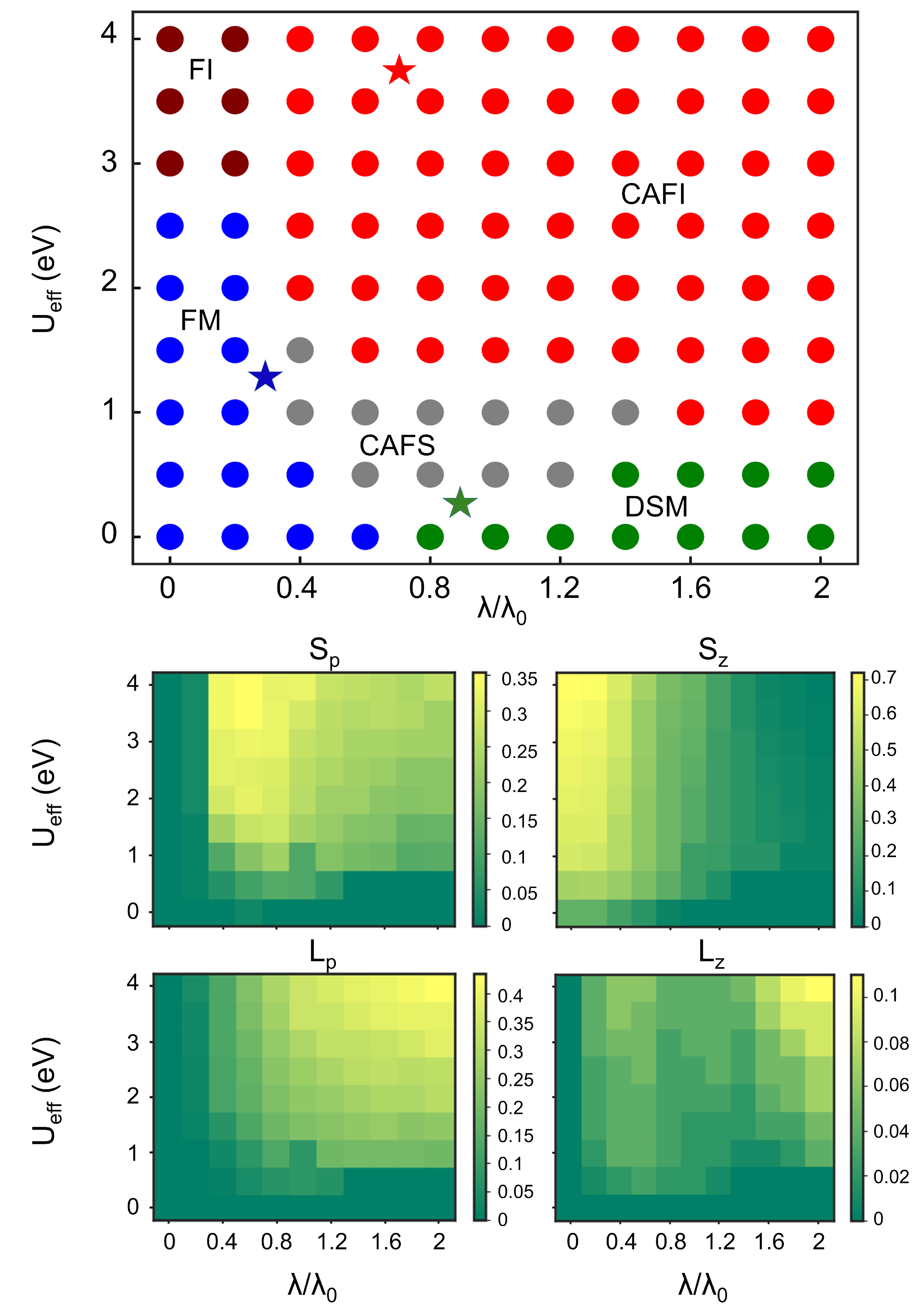}
\caption{(Top row) The electronic and magnetic phase diagram of orthorhombic SrIrO$_3$ as a function of $U_{eff}$ and SOC. (Middle and bottom rows) The planar and normal component of spin moment ($S_\mathrm{p}$ and $S_\mathrm{z}$) and orbital moment ($L_\mathrm{p}$ and $L_\mathrm{z}$) respectively. The star-marks represent the ground state of LS $d^5$ pervoskites. The green star-mark represent CaIrO$_3$, the blue star-mark represent LaRuO$_3$ and SrRhO$_3$, and the red star-mark represent YRuO$_3$, respectively.}
\label{bulk_phase}
\end{figure}

\begin{figure}[ht]
\centering
\includegraphics[angle=-0.0,origin=c,height=3.5cm,width=8cm]{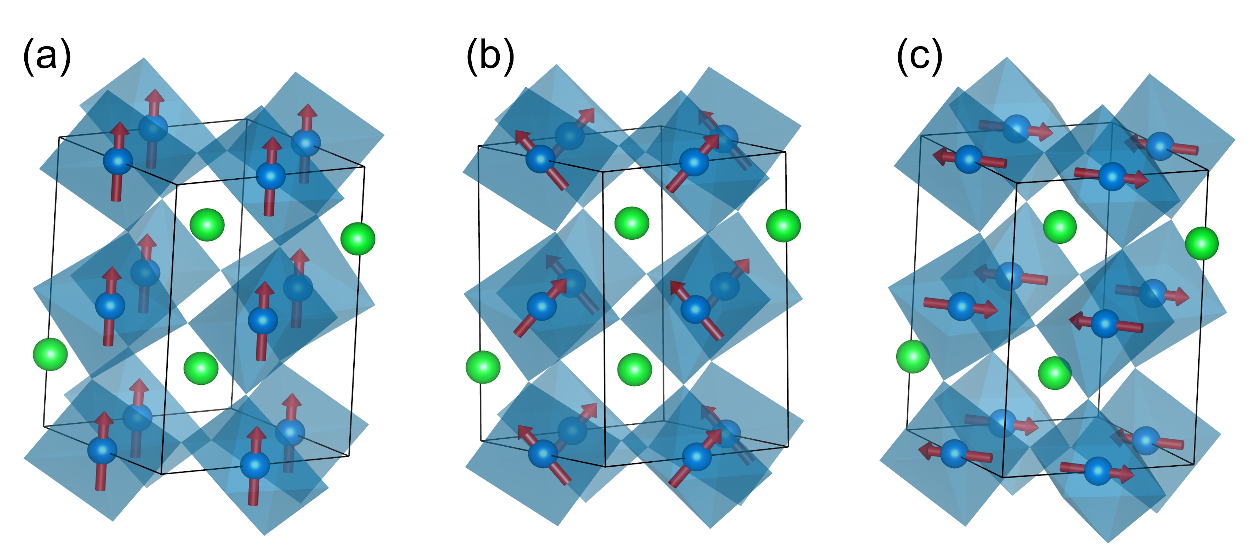}
\caption{Schematic illustrating evolution of magnetic ordering in bulk SrIrO$_3$ at $U_{eff}$ = 4 eV with (a) $\lambda$/$\lambda_0$ = 0, (b) $\lambda$/$\lambda_0$ = 1, and (c) $\lambda$/$\lambda_0$ = 2, respectively.}
\label{Fig6}
\end{figure}

The phase diagram for the orthorhombic SrIrO$_3$ spanned in $U$-SOC space, as evaluated from the DFT+$U$+SOC calculations is presented in Fig. \ref{bulk_phase}. It shows five distinct phases: (i) collinear FM as shown in blue, (ii) collinear FI as shown in maroon, (iii) canted antiferromagnetic semimetal (CAFS) as shown in grey, (iv) CAFI as shown in red, and (v) DSM as shown in green. 
Let us first examine the weak SOC regime ($\lambda$/$\lambda_o$ $\leq$ 0.2). Up to $U_{eff}$ $\approx$ $2.5$ eV, the system remains a FM with collinear magnetic ordering as shown in Fig. \ref{Fig6}(a). 
This results from the imbalance of the population of partially occupied $t_\mathrm{2g}$ states in the up-spin and down-spin channels as already explained through Figs. \ref{Fig4}(a,b) in the previous subsection. With further increase in $U$, the system undergoes a transition from FM to FI because of the fact that with enhanced spin split the earlier partially occupied band in the up-spin channel is now occupied completely whereas in the down spin channel it  becomes empty (see Fig. \ref{Fig4}(c,d)). The FI phase, as predicted from our model and DFT calculations, was not captured in the earlier study \cite{Zeb2012} where the stability of the metallic phase for all $U$ in the non-SOC regime was reported. It may be noted that the authors there have carried out the calculations using the full-potential linearized augmented-plane-wave (LAPW) method with LDA exchange-correlation functional. The strength of local magnetic moment at Ir depends on $U_{eff}$. It increases from $0.27$ $\mu_\text{B}$ at $U_{eff}$ = $0$ eV to $0.72$ $\mu_\text{B}$ at $U_{eff}$ = $4$ eV. With the LS state, the saturated magnetic moment is 1 $\mu_\text{B}$ which can be realized for further higher value of $U$. The LS state favours the ferromagnetic order over the antiferrromagnetic one. 

With finite $\lambda/\lambda_0$, the up-spin and down-spin states mix to form spin-orbit entangled pseudo-spin $J_\mathrm{eff}$ = 1/2 and $J_\mathrm{eff}$ = 3/2 states. The former occupy the $E_\mathrm{F}$ and hence determine the electronic and magnetic phase of the system. For the intermediate range (0.4 $\leq$ $\lambda$/$\lambda_0$ $\leq$ 1.4), the lower value of $U_{eff}$ maintains the metallicity and at the same time weakens the magnetic state with the formation of a DSM phase. It is important to note that the electron hopping, SOC and onsite Coulomb repulsion are in the same energy scale  and hence any perturbation can give rise to a different quantum phase. Here, we show that  even weak $U_{eff}$ transforms the DSM to \textit{CAFS} phase suggesting this weakly correlated perovskite is near to the magnetic instability. The onsite Coulomb repulsion facilitate the non-collinear ordering of the spin-orbit entangled states as the planar-spin ($S_\mathrm{p}$) component increase in magnitude in proportion to $\lambda$/$\lambda_0$ which can be observed from the spin-intensity map shown in the middle panel of Fig. \ref{bulk_phase}. The onsite repulsion also strengthens the localization to stabilize the CAFI phase as can be seen from the phase diagram. 

For the large $\lambda$/$\lambda_0$ regime ($\lambda$/$\lambda_0$ $\geq$ 1.4), the system exhibits either DSM for low $U$ or CAFI phase for lower and higher $U_{eff}$ values. In the latter case, the $S_\mathrm{z}$ component gradually vanishes with SOC (see Fig. \ref{bulk_phase} middle panel) to create a transition from the collinear along $z$ to non-coplanar and canted to pure co-planar spin arrangement. This transition is schematically illustrated in Fig. \ref{Fig6}. Earlier studies have presented an unusual trend where it has been suggested that higher value of $U_{eff}$ is required to induce metal-to-insulator transition for the higher value of SOC \cite{Zeb2012}. However, as expected, in this study we observe that a lower $U_{eff}$ value is sufficient to stabilize a CAFI phase.

The 4$d$ pervoskites SrRhO$_3$, LaRuO$_3$ and YRuO$_3$ and the $5d$ perovskites are reported to be having LS $d^5$ electonic configuration and all of them undergo GdFeO$_3$-type distortion of varied order. Their electronic and magnetic phases can be mapped to the phase diagram of Fig. \ref{bulk_phase}. The compounds SrRhO$_3$ \cite{Singh2003} and LaRuO$_3$ \cite{KOBAYASHI1994}, with weak SOC and moderate correlation exhibits FM ground state and can be placed in the region (0.2 $\leq$ $\lambda$/$\lambda_0$ $\leq$ 0.4, 1 $\leq$ $U_{eff}$ $\leq$ 2 eV). The compound YRuO$_3$ \cite{Ji2020} exhibits CAFI phase and lies in (0.6 $\leq$ $\lambda$/$\lambda_0$ $\leq$ 0.8, 3 $\leq$ $U_{eff}$ $\leq$ 4 eV) zone. Like SrIrO$_3$, CaIrO$_3$ \cite{Fujioka2019} stabilizes in nonmagnetic DSM phase and lies in  (0.8 $\leq$ $\lambda$/$\lambda_0$ $\leq$ 1, 0 $\leq$ $U_{eff}$ $\leq$ 0.5 eV) zone of the phase diagram. In Appendix C, we have compared the phase diagrams of SrIrO$_3$ and CaIrO$_3$ (see Fig. 11) to demonstrate the generality of the electronic structure evolving out of competition between electron-electron correlation and SOC.
\begin{figure}[hbt]
\centering
\includegraphics[angle=-0.0,origin=c,height=7cm,width=8.5cm]{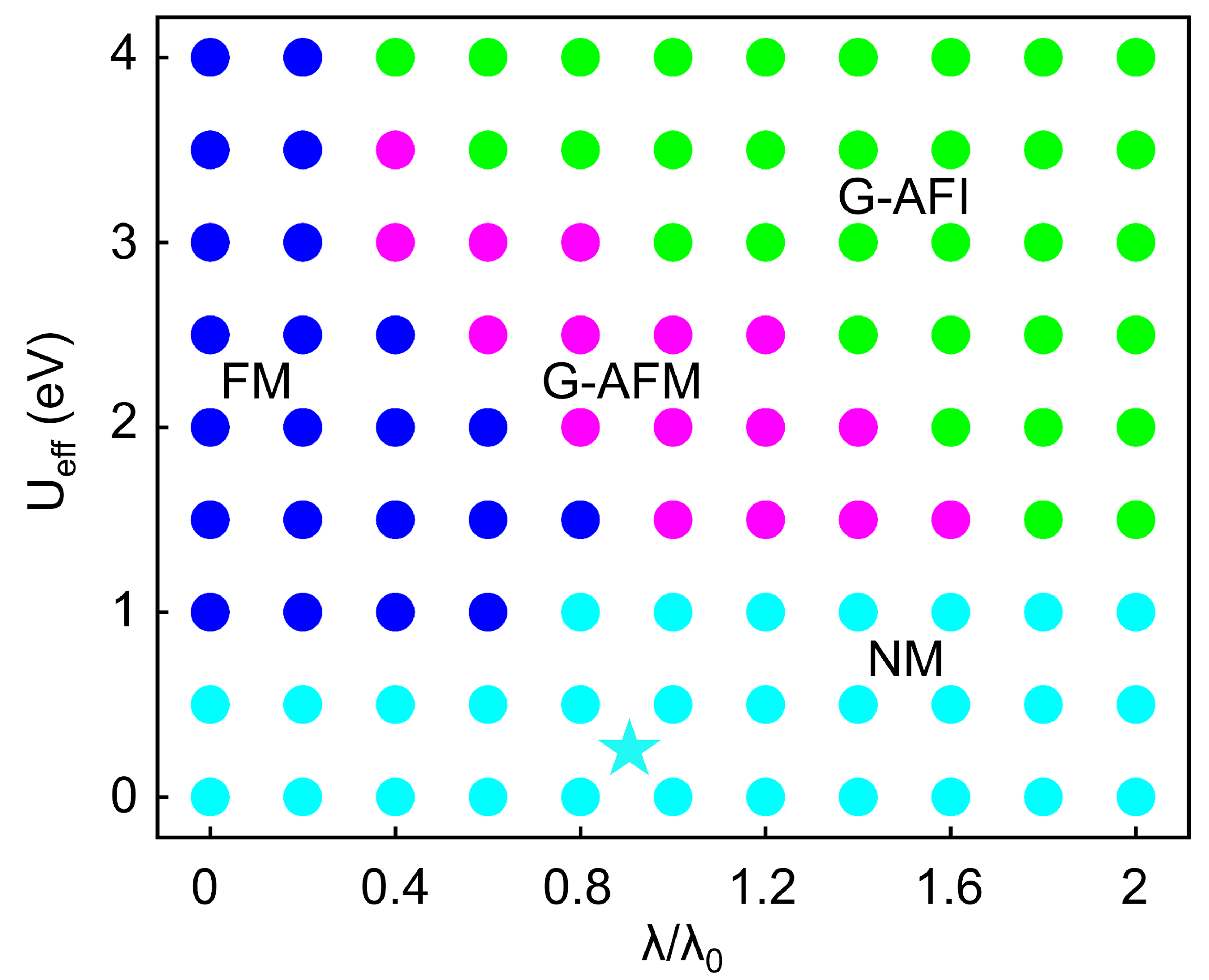}
\caption{The electronic and magnetic phase diagram of undistorted (cubic) SrIrO$_3$ as a function of $U_{eff}$ and SOC. The star-mark with cyan color represents the ground state of LS $d^5$ pervoskite BaIrO$_3$, respectively}
\label{cubic_phase}
\end{figure}

To identify the divison of role between SOC and structural distortions in estabilishing quantum phases, here, we have computed the phase diagram for undistorted (cubic) SrIrO$_3$ in the $U$-SOC space (see Fig. \ref{cubic_phase}). Most significantly, we observe that the non-collinear spin-ordering is missing and the entire phase diagram is spanned by collinear magnetic phases. 
There are four distinct phases observed and these are: (i) FM, as shown in blue (ii) G-AFM, as shown in magenta, (iii) G-AFI, as shown in lime and (iv) NM, as shown in cyan. In the weak SOC ($\lambda$/$\lambda_0$ $\leq$ 0.2) and weak $U_{eff}$ regime, the system stabilizes in the NM state contrary to the FM phase, driven by small but finite moment, in the distorted structure. As the strength of onsite Coloumb repulsion increases, unlike the case of distorted structure, here no phase transition occurs and the system remains in the FM phase. The FM phase, for higher $U_{eff}$ values is well captured by our model Hamiltonian for the undistorted structure. In Appendix D, we have analyzed the orbital and spin resolved density of states (DOS) to show the robustness of the metallic phase for higher values of $U_{eff}$.
For the intermediate SOC strength (0.4 $\leq$ $\lambda$/$\lambda_0$ $\leq$ 1.6), there is a narrow domain in which G-AFM phase stabilizes. In the strong $U$ limit, transition from G-AFM to G-AFI phase occurs where the $U_\mathrm{c}$ value for metal-insulator transition is found to be high as compared to the distorted case. For example, $U_\mathrm{c}$ varies from 1 to 2 eV, whereas, for the undistorted case, a higher $U_\mathrm{c}$ between 2 to 4 eV is required for metal-insulator transition. The bandwidth, a measure of extent of localization, in a Hubbard model weakens the hopping integral roughly by a factor of 1/$U$. The distortion also weakens the hopping integral. Therefore, metal-insulator transition can be achieved with a lower value of $U$ for the distorted structure. For the large SOC domain ($\lambda$/$\lambda_0$ $\geq$ 1.6), system exhibits either NM phase for lower $U_{eff}$ values or G-AFI phase for intermediate and higher $U_{eff}$ values. The compound BaIrO$_3$ \cite{Cheng2013}, like SrIrO$_3$, exhibits Pauli paramagnetic ground state and can be mapped to the region (0.8 $\leq$ $\lambda$/$\lambda_0$ $\leq$ 1, 0 $\leq$ $U$ $\leq$ 1 eV) in the phase diagram.

\section{Origin of Non-Collinear Magnetism: Effect of Rotation and Tilting}
\begin{figure}[hbt]
\centering
\includegraphics[angle=-0.0,origin=c,height=12.5cm,width=8.5cm]{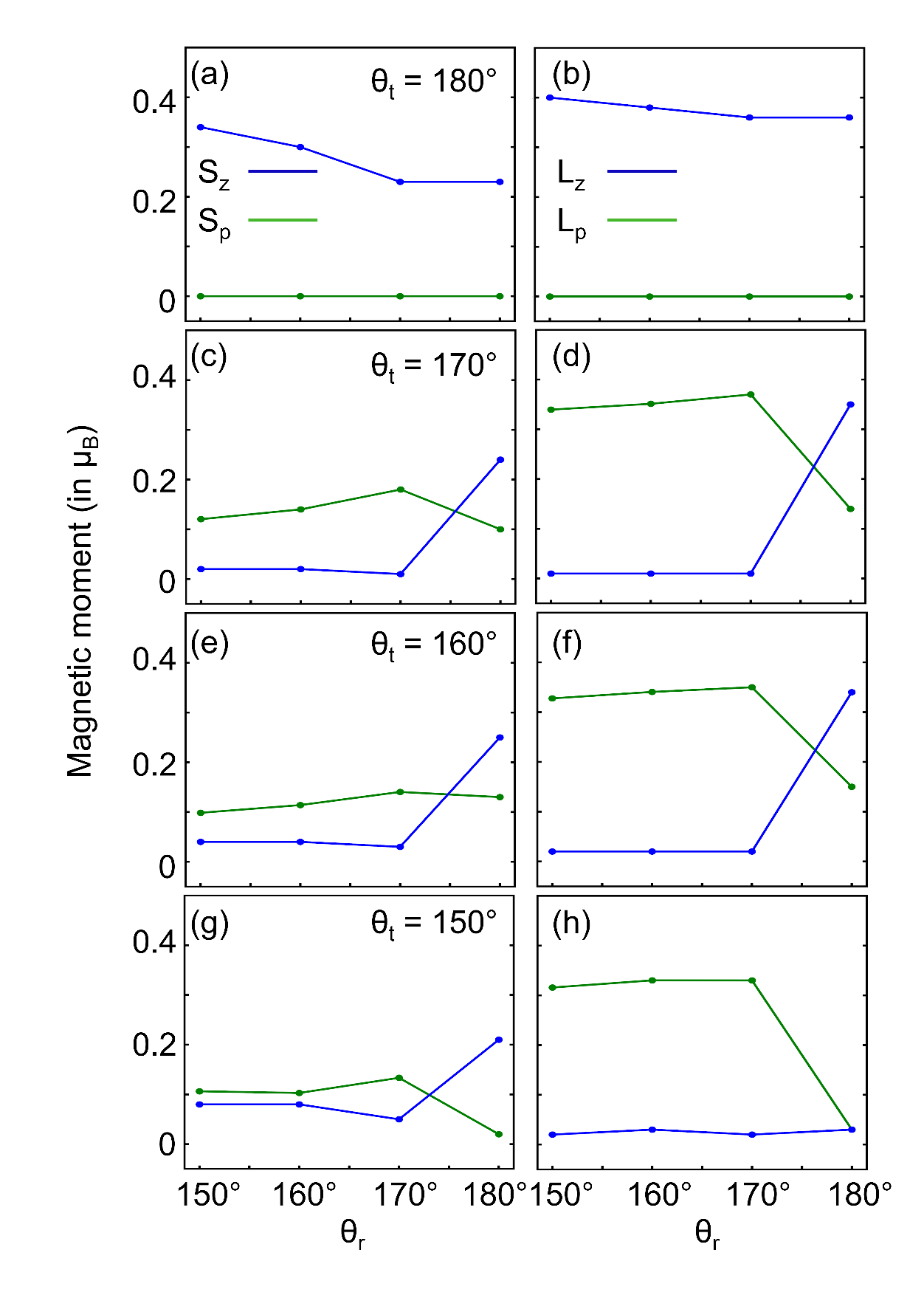}
\caption{Variation of planar and normal spin (first coloumn) and orbital moments (second coloumn) with rotation and tilting of octahedra, respectively.}
\label{spin_orbital_mom}
\end{figure}
Absence of planar-spin component in the cubic SrIrO$_3$ implies that distortion is the key to the stabilization of non-collinear spin ordering. The effect of distortions on magnetic ordering is also observed in iridates and their superlattices. For example, Sr$_3$Ir$_2$O$_7$ exhibits a robust c-axis collinear antiferromagnetic ordering with negligible $\theta_\mathrm{t}$ = 179.5$^\circ$ \cite{Hogan2016,Fujiyama2012} but on the other side, the bilayer superlattice 2SIO/1STO, exhibits c-axis canted AF ordering which is attributed to the presence of finite $\theta_\mathrm{t}$ ($\approx$ $172^\circ$). It has been reported that further enhancement of $\theta_\mathrm{t}$ beyond 172$^\circ$ can even drive the 2SIO/1STO through a quantum critical point where out-of-plane collinear to in-plane canted magnetic phase transition occurs \cite{Meyers2019}. Therefore, to analyze the effect of structural distortions on the LS state of $d^5$ pervoskites, we have provided a quantitative measure of it by creating a geometrical design, described in Appendix \ref{B} where the $\theta_\mathrm{r}$ and $\theta_\mathrm{t}$ can be varied smoothly.

For $\lambda$/$\lambda_0$ = 1 and $U_{eff}$ = 4 eV, the planar and normal spin and orbital moments are plotted as a function of $\theta_\mathrm{r}$ and $\theta_\mathrm{t}$ in Fig. \ref{spin_orbital_mom}. In the absence of tilting, $\theta_\mathrm{t}$ = 180$^\circ$, see Figs. \ref{spin_orbital_mom}(a,b), system always favours collinear spin ordering with vanishing $S_\mathrm{p}$ and $L_\mathrm{p}$ components. As far as $S_\mathrm{z}$ is concerned, it increases with $\theta_\mathrm{r}$ for $\theta_\mathrm{t}$ = 180$^\circ$, hence, increasing the total magnetization of the system. Finite tilting (see Figs. \ref{spin_orbital_mom}(c,e,g)) leads to non-coplanar spin arrangement where $S_\mathrm{z}$ decreases sharply with increasing $S_\mathrm{p}$ component upto $\theta_\mathrm{r}$ = 170$^\circ$. With further increase in $\theta_\mathrm{r}$, $S_\mathrm{p}$ component decreases whereas $S_\mathrm{z}$ component increases and finally both components become comparable for higher values of $\theta_\mathrm{r}$. More or less the $L_\mathrm{p}$ component also follow the same trend as the $S_\mathrm{p}$ with increase in $\theta_\mathrm{r}$ but with magnitude larger as compare to the $S_\mathrm{p}$ (see Figs. \ref{spin_orbital_mom}(d,f,h)). However, for $\theta_\mathrm{t}$ = 150$^\circ$, the $L_\mathrm{z}$ component vanishes completely suggesting quenching of $L_\mathrm{z}$ component for higher $\theta_\mathrm{t}$. The isospin reorientation with tilting and rotation that we get from this first-principles study is attributed to the orbital mixing hoppings arising in the distorted frame. Many body models are being designed to relate such spin anisotropy to the Kitaev-type interactions developed due to octahedral tilting \cite{Mohapatra2019}. 

\section{Summary and Outlook}
To summarize, we developed a multi-band model Hamiltonian and performed density functional calculations to study the electronic and magnetic structure of SrIrO$_3$ and examined the role of onsite Coulomb repulsion and spin-orbit coupling in the cubic and distorted structural framework. Furthermore, we used SrIrO$_3$ as a prototype to examine the electronic structure of low-spin $d^5$ perovskites in general by building phase diagram and also smoothly varied the rotation and tilting of the IrO$_6$ octahedra to bring a third dimension into it. Our study reveals that eight quantum phases, namely, nonmagnetic metal, nonmagnetic Dirac semimetal, ferromagnetic metal, ferromagnetic insulator, G-type antiferromagnetic metal, G-type antiferromagnetic insulator, canted antiferromagnetic semimetal, and canted antiferromagnetic insulator as shown in Figs. \ref{bulk_phase} and \ref{cubic_phase} of the main text. The mechanism driving such phases are explained in details. Further, we find that each of them form a soft-boundary to allow a continuous phase transition from one phase to the other by varying the interaction strengths. The phase diagram is validated by mapping the ground state of the reported low-spin $d^5$ perovskites CaIrO$_3$, BaIrO$_3$, SrRhO$_3$, LaRuO$_3$ and YRuO$_3$ in the phase diagrams.

By scanning the periodic table, we see that low-spin $d^5$ transition metal oxide perovskites can be designed by exploring the following group combinations: I-X, II-IX, III-VIII (KPdO$_3$, RbPdO$_3$, MgRhO$_3$, ScRuO$_3$, ScOsO$_3$ and YOsO$_3$, etc.). Theoretically, thermodynamical stability of such systems can be examined and experimental synthesis can be attempted with the advent of state of the art synthesis techniques such as atom by atom deposition methods and  high pressure methods. Furthermore, the two formula unit double perovskite transition metal oxides with $d^5$ state can be thought of as sister members where similar competing interactions govern the system. In this way, the phases proposed in the phase diagram can be achieved. The interaction strengths can be varied under external stimuli such as pressure and strain, as well as through changing the chemical composition, design of heterostructures, etc. to induce quantum phase transition in these systems. As a whole, we believe that the present study will trigger experimental and theoretical studies to envisage novel quantum phases and applications.

\section{Acknowledgement}
The authors would like to thank HPCE, IIT Madras for providing the computational facility. This work is funded by the Department of Science and Technology, India, through grant No. CRG/2020/004330.

\appendix
\section{TRANSFORMATION OF HOPPING MATRICES UNDER ROTATION AND TILTING OF OCTAHEDRA}
\label{A}
In this appendix we briefly explain the transformation of TB matrices for hopping between different sublattices under rotation and tilting of octahedra. Distortion can be described in the form of octahedral rotation and tilting which varies from one Ir-site to the another as shown in Fig. \ref{Fig1}(a). Here, $+$ or $-$ signs on each Ir-site indicates the clockwise or counterclockwise rotation and tilting of octahedra. The transformation of hopping matrices is determined by a site dependent $5$ $\times$ $5$ rotation matrix (Eq. A4) \cite{Mohapatra2018,Tinkham}. Denoting unrotated and rotated basis in the order (xy,yz,xz,x$^2$ - y$^2$,3z$^2$ - r$^2$) for A sublattice as $\ket{\alpha}$ and $\ket{\alpha^\prime}$, for B sublattice as $\ket{\beta}$ and $\ket{\beta^\prime}$, so that $\ket{\alpha^\prime}$ = R $\ket{\alpha}$ and $\ket{\beta^\prime}$ = R$^\prime$ $\ket{\beta}$, where R and R$^\prime$ are corresponding rotation matrices for A and B sublattices, the hopping integral in the rotated basis is then given by,
\begin{equation}
       \tilde{H}_{\alpha^\prime,\beta^\prime}= \bra{\alpha^\prime} H \ket{\beta^\prime}\\
         = \bra{\alpha} R^T H R^\prime \ket{\beta} 
\end{equation}
Hence, the Hamiltonian in the rotated basis is given by,
\begin{equation}
    \tilde{H} = R^T H R^\prime \label{eqA1}
\end{equation}
The total rotation matrix for each site for Ir site can be described as the multiplication of two individual rotation and tilt matrices. The product is given by,
\begin{equation}
    R(\theta_\mathrm{r},\theta_\mathrm{t}) = R(\theta_\mathrm{r}) R(\theta_\mathrm{t})
\end{equation}
where R($\theta_\mathrm{r}$) denotes the pure rotation about the z axis whereas R($\theta_\mathrm{t}$) denotes the rotation about the crystal axis $a$. For pure rotation about z axis, the Euler angles ($\alpha$,$\beta$,$\gamma$) = ($\theta_\mathrm{r}$,0,0) whereas for tilting ($\alpha$,$\beta$,$\gamma$) = (-45$^\circ$,$\theta_\mathrm{t}$,45$^\circ$), respectively. Using these Euler angles, R($\theta_\mathrm{r}$) and R($\theta_\mathrm{t}$) can be obtained using R given by
\begin{widetext}
\begin{equation}
    R = \frac{1}{2}\begin{pmatrix}
       \beta{^\prime_+}^2 \cos(2\alpha+2\gamma)/2 &  \beta{^\prime_+}\sin\beta \cos(2\alpha+\gamma) & \beta{^\prime_+}\sin\beta \sin(2\alpha+\gamma)  & \beta{^\prime_+}^2 \sin(2\alpha+2\gamma)/2  & \sqrt{3}\sin^2\beta \\[0.5em]
       
       -\beta{^\prime_-}^2 \cos(2\alpha-2\gamma)/2 & +\beta{^\prime_-}\sin\beta \cos(2\alpha-\gamma) & -\beta{^\prime_-}\sin\beta \sin(2\alpha-\gamma) & +\beta{^\prime_-}^2 \sin(2\alpha-2\gamma)/2 & \sin{2\alpha}\\[2em]
       
        -\beta{^\prime_+}\sin\beta\cos(\alpha+2\gamma) &  \beta{^\prime_+}\beta^{\prime\prime}_- \cos(\alpha+\gamma) & \beta{^\prime_+}\beta^{\prime\prime}_- \sin(\alpha+\gamma)  & -\beta{^\prime_+}\sin\beta\sin(\alpha + 2\gamma)  & \sqrt{3}\sin\alpha \\
       
       -\beta{^\prime_-} \sin\beta\cos(\alpha-2\gamma) & +\beta{^\prime_-}\beta^{\prime\prime}_+ \cos(\alpha-\gamma) & -\beta{^\prime_-}\beta^{\prime\prime}_+ \sin(\alpha-\gamma) & +\beta{^\prime_-} \sin\beta\sin(\alpha-2\gamma)& \sin{2\beta}\\[2em]
       
       \beta{^\prime_+}\sin\beta\sin(\alpha+2\gamma) &  -\beta{^\prime_+}\beta^{\prime\prime}_- \sin(\alpha+\gamma) & \beta{^\prime_+}\beta^{\prime\prime}_- \cos(\alpha+\gamma)  & -\beta{^\prime_+}\sin\beta\cos(\alpha + 2\gamma)  & \sqrt{3}\cos\alpha \\
       
       +\beta{^\prime_-} \sin\beta\sin(\alpha-2\gamma) & -\beta{^\prime_-}\beta^{\prime\prime}_+ \sin(\alpha-\gamma) & -\beta{^\prime_-}\beta^{\prime\prime}_+ \cos(\alpha-\gamma) & +\beta{^\prime_-} \sin\beta\cos(\alpha-2\gamma)& \sin{2\beta}\\[2em]
       
        -\beta{^\prime_+}^2 \sin(2\alpha+2\gamma)/2 &  -\beta{^\prime_+}\sin\beta \sin(2\alpha+\gamma) & \beta{^\prime_+}\sin\beta \cos(2\alpha+\gamma)  & \beta{^\prime_+}^2 \cos(2\alpha+2\gamma)/2  & \sqrt{3}\sin^2\beta \\[0.5em]
       
       +\beta{^\prime_-}^2 \sin(2\alpha-2\gamma)/2 & -\beta{^\prime_-}\sin\beta \sin(2\alpha-\gamma) & -\beta{^\prime_-}\sin\beta \cos(2\alpha-\gamma) & \beta{^\prime_-}^2 \cos(2\alpha-2\gamma)/2 & \cos{2\alpha}\\[2em]
       
       -\sqrt{3}\sin^2\beta\sin2\gamma & \sqrt{3}\sin2\beta\sin\gamma & - \sqrt{3}\sin2\beta\cos\gamma & \sqrt{3}\sin^2\beta\cos2\gamma & 2-3\sin^2\beta
      
       \end{pmatrix} 
       \end{equation}
\end{widetext}
where $\beta_\pm^\prime$ = 1 $\pm$ $\cos\beta$ and $\beta_\pm^{\prime\prime}$ = 2 $\cos\beta$ $\pm$ 1.

In the rotated basis, the TB hopping matrices between A to B sublattice can be obtained by using the following equation,
\begin{equation}
    \tilde{H}_\mathrm{AB} = R^T(-\theta_\mathrm{r},-\theta_\mathrm{t})H_\mathrm{AB} R(\theta_\mathrm{r},-\theta_\mathrm{t})
\end{equation}
where $H_\mathrm{AB}$ is the Hamiltonian is the unrotated basis. In compact form $H_\mathrm{AB}$ is expressed as,
\begin{equation}
H_\mathrm{AB}= \begin{pmatrix}
       H_\mathrm{AB}^{\uparrow \uparrow}& H_\mathrm{AB}^{\uparrow \downarrow}\\[0.2cm]
       H_\mathrm{AB}^{\downarrow \uparrow } & H_\mathrm{AB}^{\downarrow \downarrow}
       \end{pmatrix}     
\end{equation}

with H$^{\uparrow\uparrow}_\mathrm{AB}$ =H$^{\downarrow\downarrow}_\mathrm{AB}$ and H$^{\uparrow\downarrow}_\mathrm{AB}$ = H$^{\downarrow\uparrow}_\mathrm{AB}$ = 0. The submatrix H$_\mathrm{AB}^{\uparrow\uparrow}$ in the expanded form is given by,
\begin{equation}
      H_\mathrm{AB}^{\uparrow\uparrow} = \begin{pmatrix}
       k_1F_1 & 0 & 0 & 0 & 0  \\[0.2cm]
       0 & k_2F_1 &  k_3F_2 & 0 & 0 \\[0.2cm]
       0 & k_3F_2 & k_2F_1 & 0 & 0 \\[0.2cm]
       0 & 0 & 0 & k_4F_1 & k_5F_2  \\[0.2cm]
       0 & 0 & 0 & k_5F_2 & k_6F_1  \\[0.2cm]
       \end{pmatrix} 
\end{equation}

where,
\begin{align}
 F_1 &= \cos(k_x/2)\cos(k_y/2)\notag\\ 
 F_2 &= \sin(k_x/2)\sin(k_y/2)\notag\\
 k_1 &= 4 t_1\notag\\
 k_2 &= 2 t_1\notag\\
 k_3 &= -2 t_1\notag\\
 k_4 &= 3 t_4\notag\\
 k_5 &= \sqrt{3} t_4\notag\\
 k_6 &= t_4 
 \end{align}
 
Similarly, Hopping between A to A sublattice is given by,
\begin{equation}
    \tilde{H}_\mathrm{AA} = R^T(-\theta_\mathrm{r},-\theta_\mathrm{t})H_\mathrm{AA} R(-\theta_\mathrm{r},-\theta_\mathrm{t})\label{eq:A10}
\end{equation} 
where,
 \begin{equation}
      H_\mathrm{AA}^{\uparrow\uparrow} = \begin{pmatrix}
       k_7F_3 & 0 & 0 & 0 & k_8F_4  \\[0.2cm]
       0 & k_{9}F_6 & 0 & 0 & 0 \\[0.2cm]
       0 & 0 & k_{9}F_5 & 0 & 0 \\[0.2cm]
       0 & 0 & 0 & k_{10}F_3 & 0  \\[0.2cm]
       k_8F_4 & 0 & 0 & 0 & k_{11}F_3 
       \end{pmatrix} 
\end{equation}
and the corresponding dispersions and hopping strength are given as,
\begin{align}
 F_3 &= \cos{k_x} + \cos{k_y}\notag\\
 F_4 &= \cos{k_x} - \cos{k_y}\notag\\
 F_5 &= \cos{k_x}\notag\\
 F_6 &=  \cos{k_y}\notag\\
 k_7 &= \frac{3}{2}t_2\notag\\
 k_8 &= -\frac{\sqrt{3}}{2}t_7\notag\\
 k_9 &= 2 t_3 \notag\\
 k_{10} &= 2 t_6 \notag\\
 k_{11} &= \frac{1}{2} t_5
 \end{align}
 
Similarly, the transformation and corresponding sublattice hopping matrices for interlayer coupling are given by,
\begin{align}
    \tilde{H}_\mathrm{AC} &= R^T(-\theta_\mathrm{r},-\theta_\mathrm{t})H_\mathrm{AC} R(-\theta_\mathrm{r},\theta_\mathrm{t})\\
     \tilde{H}_\mathrm{AD} &= R^T(-\theta_\mathrm{r},-\theta_\mathrm{t})H_\mathrm{AD} R(\theta_\mathrm{r},\theta_\mathrm{t})
\end{align}

\begin{align}
      H_\mathrm{AC}^{\uparrow\uparrow} &= \begin{pmatrix}
       0 & 0 & 0 & 0 & 0  \\[0.2cm]
       0 & k_{12}F_7 & 0 & 0 & 0 \\[0.2cm]
       0 & 0 & k_{12}F_7 & 0 & 0 \\[0.2cm]
       0 & 0 & 0 & 0 & 0  \\[0.2cm]
       0 & 0 & 0 & 0 & k_{13}F_7
       \end{pmatrix}
       \end{align}
      
      \begin{align}\label{eq:A16}
       H_\mathrm{AD}^{\uparrow\uparrow} &= \begin{pmatrix}
       k_{14}F_8 & k_{15}F_{11} &  k_{15}F_{10} & 0 & 0  \\[0.2cm]
       k_{15}F_{11} & k_{16}F_{8} & k_{17}F_{9}&  k_{18}F_{10} & k_{19}F_{11} \\[0.2cm]
       k_{15}F_{10} &  k_{17}F_{9} & k_{16}F_8 &  k_{18}F_{11} &  k_{19}F_{10} \\[0.2cm]
       0 & k_{18}F_{10} &  k_{18}F_{11} &  k_{20}F_{8} &  k_{21}F_{9}  \\[0.2cm]
       0 &  k_{19}F_{11} & k_{19}F_{10} &  k_{21}F_{9} &  k_{22}F_{8} 
       \end{pmatrix} 
\end{align}
and the corresponding dispersion relations and hopping strength are given as,

\begin{align}
 F_7 &= \cos(k_z/2)\notag\\
 F_8 &= \cos(k_x/2)\cos(k_y/2)\cos(k_z/2)\notag\\
 F_9 &= \sin(k_x/2)\sin(k_y/2)\cos(k_z/2)\notag\\
 F_{10} &= \sin(k_x/2)\sin(k_z/2)\cos(k_y/2)\notag\\
 F_{11} &= \sin(k_y/2)\sin(k_z/2)\cos(k_x/2)\notag\\
 k_{12} &= 2t_1\notag\\
 k_{13} &= 2t_4\notag\\
 k_{14} &=  4t_3\notag\\
 k_{15} &=  -2\sqrt{2}t_3\notag\\
 k_{16} &=  (3t_2 + 2t_3)\notag\\
 k_{17} &=  -(3t_2 - 2t_3)\notag\\
 k_{18} &=  -\frac{3}{\sqrt{2}}t_7\notag\\
 k_{19} &=  -\sqrt{\frac{3}{2}}t_7\notag\\
 k_{20} &=  (\frac{3}{2}t_5 + 2t_6)\notag\\
 k_{21} &=  (-\sqrt{3}t_5 + 2t_6) \notag\\
 k_{22} &=  (\frac{t_5}{2} + 6t_6)
 \end{align}
 
\section{GEOMETRICAL DESIGN FOR INDUCING ROTATION AND TILTING}
\label{B}

\begin{figure}[hbt]
\centering
\includegraphics[angle=-0.0,origin=c,height=6.5cm,width=8.5cm]{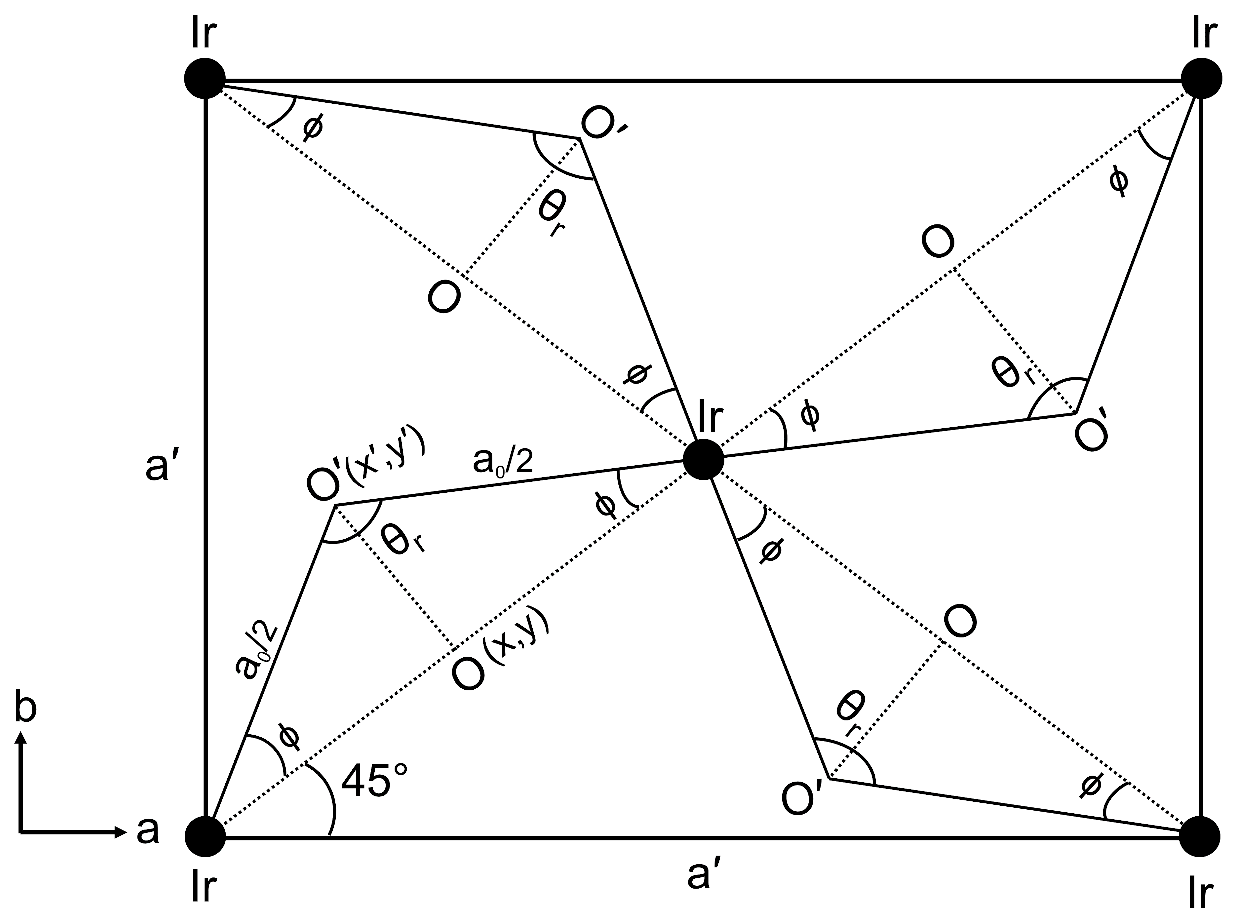}
\caption{Schematic illustration of geometrical design for inducing rotation in the $ab$ plane. Here, a (= $\sqrt{2} a_\mathrm{0}$) and a$^\prime$ are the lattice parameters of the undistorted and distorted structures, respectively.}
\label{rot}
\end{figure}

\begin{figure}[hbt]
\centering
\includegraphics[angle=-0.0,origin=c,height=8cm,width=8.5cm]{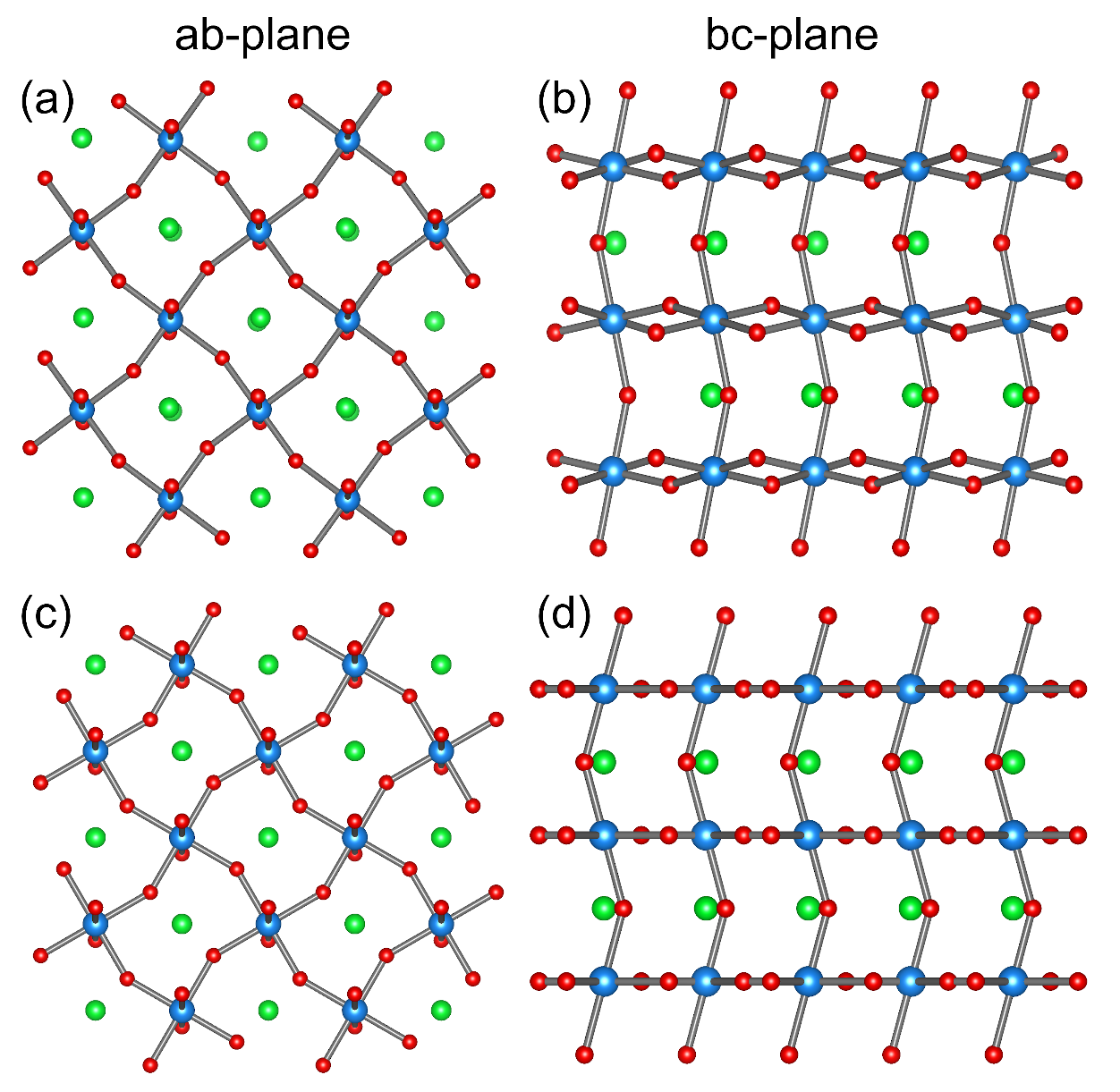}
\caption{Planar view of $ab$ and $bc$ planes for (a,b) real orthorhombic structure, and (c,d) for the designed supercell geometry.}
\label{dis_struct}
\end{figure}

The geometrical design for inducing rotation in the $ab$ plane is shown in Fig. \ref{rot}. This geometrical design differs from the equilibrium orthorhombic structure (see Fig. \ref{dis_struct}). For the former, the Ir-O bond lengths are uniform while the Ir-O-Ir bond-angles vary to facilitate the rotation and tilting. However, for the latter both bond-lengths and bond angles are anisotropic to minimize the energy. Therefore, this geometrical design can explicitly examine the effect of rotation and tilting on spin ordering. To do a quantitative measure of it, we have specifically considered the case of $bc$ plane tilting only.
For analyzing the effect of structural distortions, a $\sqrt{2}a_\mathrm{0} \times \sqrt{2}a_\mathrm{0} \times 2a_\mathrm{0}$ supercell is designed. Further distortion is induced in the supercell geometry by displacing the oxygen atoms in the $ab$ (for rotation) as well in the $bc$ (for tilting) plane keeping the  Ir-O bond length fixed. The geometrical design for inducing rotation is shown in Fig. \ref{rot}. For facilitating $\theta_\mathrm{r}$, oxygen atoms are displaced from their mean position O(x,y) to O$^\prime$(x$^\prime$, y$^\prime$). In this process the lattice parameter changes from $a$ to a$^\prime$ to maintain the constant bond length. By using similar geometry the tilting is designed in the $bc$ plane. The lattice parameter for distorted structure and the coordinates of displaced oxygen atoms are related to each other by the following equations,
\begin{align}
    x^\prime &= \frac{a^\prime \cos{(\phi + 45^\circ)}}{2\sqrt{2}\cos{\phi}}\\
    y^\prime &= \frac{a^\prime \sin{(\phi + 45^\circ)}}{2\sqrt{2}\cos{\phi}}\\
    a^\prime &= a\cos{\phi}
\end{align}
\section{Comparison between bulk CaIrO$_3$ and SrIrO$_3$}
We have carried out calculations using the experimentally synthesized structure of CaIrO$_3$ \cite{Tsuchiya2007} to examine the possible phases that this compound exhibits in the $U_{eff}$-$\lambda$ space. From the structural point of view CaIrO$_3$ and SrIrO$_3$ differs largely through tilting and rotation angles ($\theta_r$ and $\theta_t$). For the former these are 153$^\circ$ and 156$^\circ$ while for the latter these are 141$^\circ$  and 143$^\circ$.\par
\begin{figure}[H]
\centering
\includegraphics[angle=-0.0,origin=c,height=9cm,width=9cm]{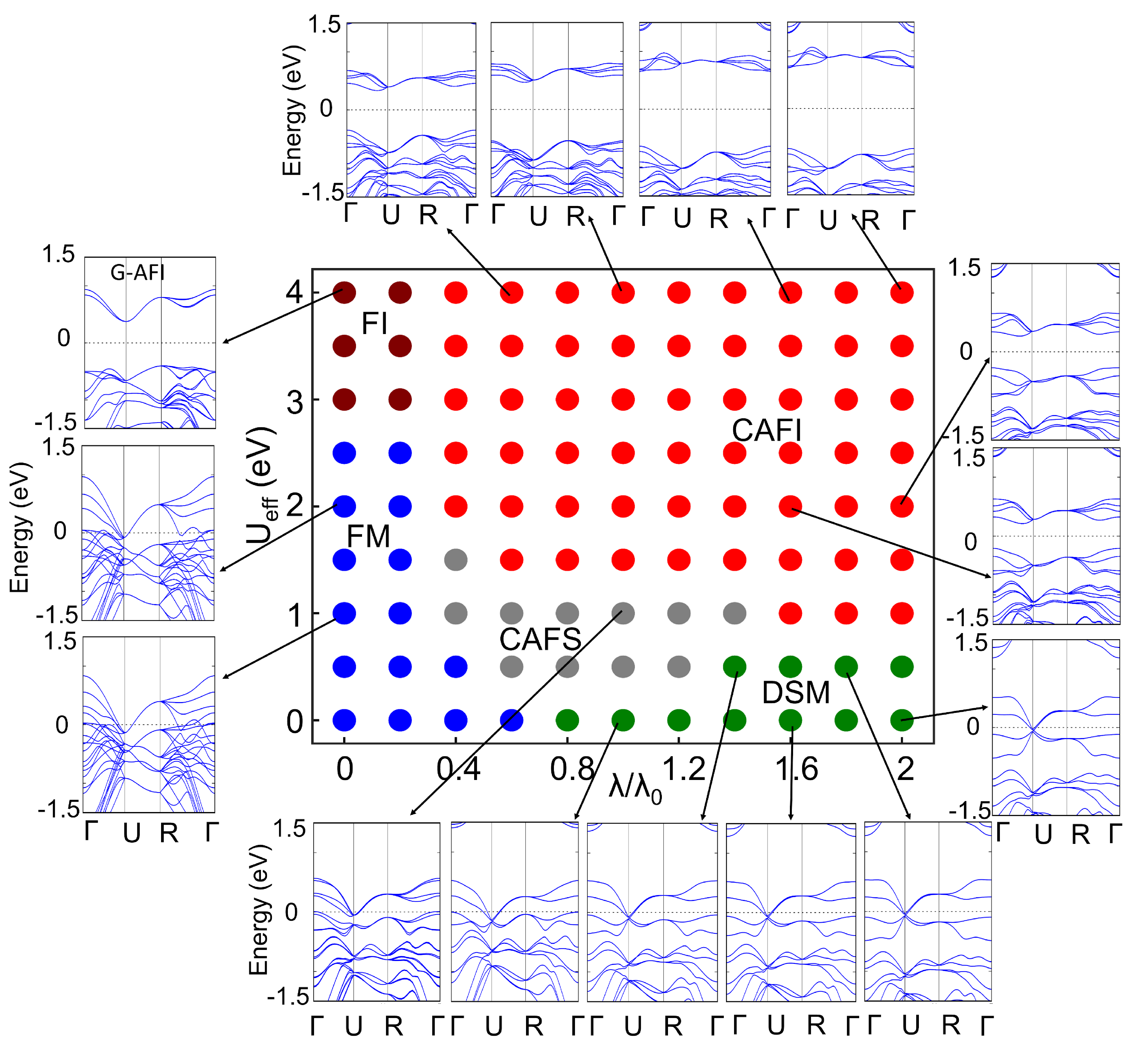}
\caption{Electronic band structures of CaIrO$_3$ marked on bulk phase diagram of SrIrO$_3$ for different values of $U_{eff}$ and $\lambda$/$\lambda_0$.}
\label{CIO_bands}
\end{figure}
\begin{figure}[H]
\centering
\includegraphics[angle=-0.0,origin=c,height=6cm,width=6.5cm]{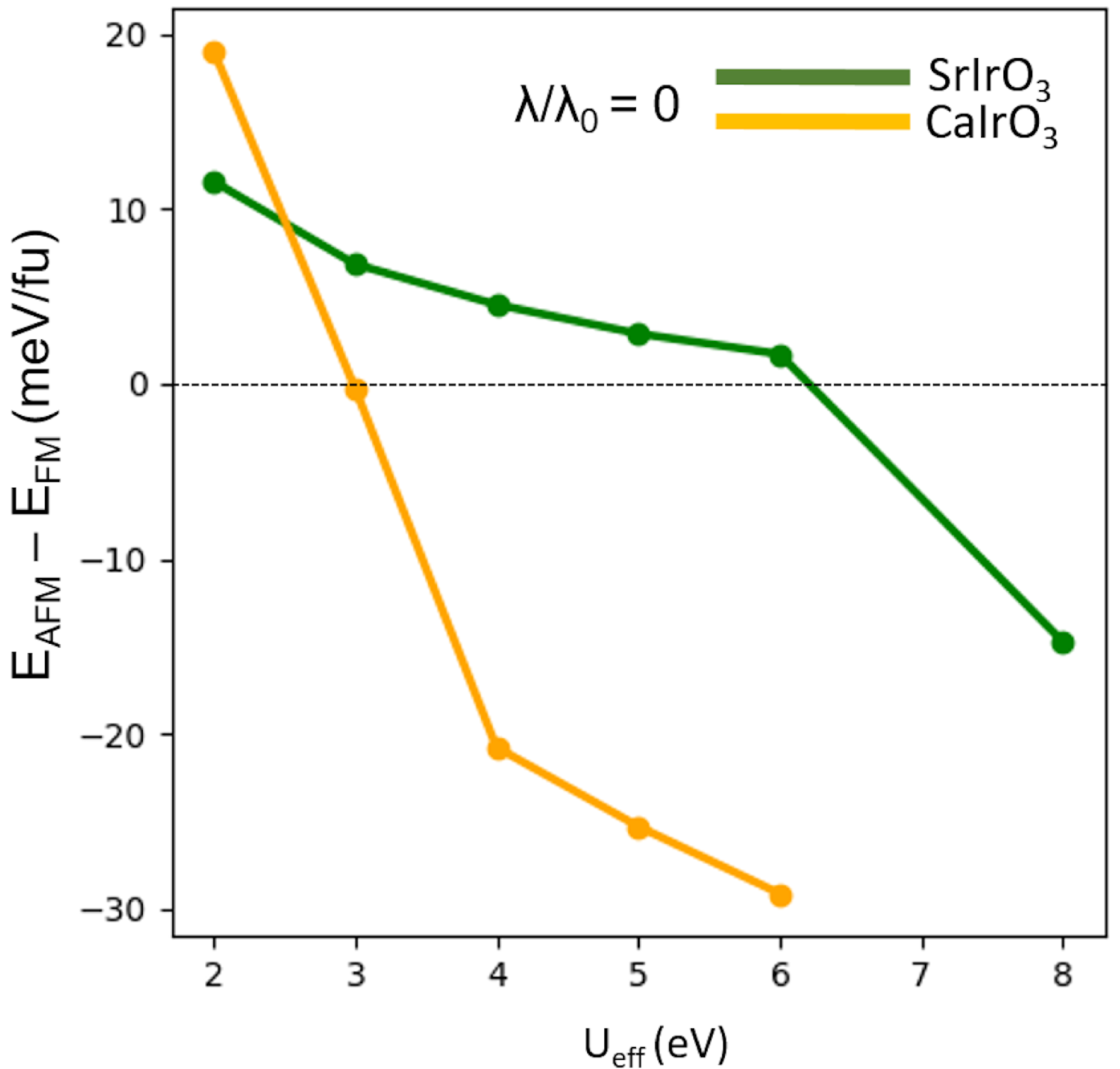}
\caption{Energy difference between G-type antiferromagnetic and ferromagnetic states as a function of $U_{eff}$ for experimentaly synthesized structures of SrIrO$_3$ and CaIrO$_3$ at $\lambda$/$\lambda_0$ = 0.}
\label{Fig11}
\end{figure}

For comparison, instead of replicating the full phase diagram of SIO (see Fig. 5) we have picked multiple points from each domain and calculated the electronic structure on these points to validate the acceptability of the phase diagram. The results are shown in Fig. \ref{CIO_bands}.
We indeed found the stabilization of five phases: FM, G-AFI, CAFS, CAFI and DSM. The electronic structure of few selective points are marked on the bulk phase diagram of SIO and are shown in Fig. \ref{CIO_bands}. The ground state DSM phase of CIO is very well captured (see band structure in the bottom pannel). Also, as seen in the case of SIO, the CAFI phase stabilizes for intermediate and higher values of $\lambda$/$\lambda_0$ and $U_{eff}$.
Here, the boundaries of the phase diagrams should not be treated as hard boundaries and they might vary depending on the compound and due to different rotation and tilting angles. As can be seen from Fig. \ref{Fig11}, in the case of CIO, we find that the system stabilizes in the G-AFI state beyond $U_{eff}$ = 3 eV whereas for SIO the G-AFI state stabilizes beyond 6 eV. These G-AFI states are formed with low-spin $d^5$ states.
\section{Robustness of the cubic metallic phase}
\begin{figure}[H]
\centering
\includegraphics[angle=-0.0,origin=c,height=9.5cm,width=6.3cm]{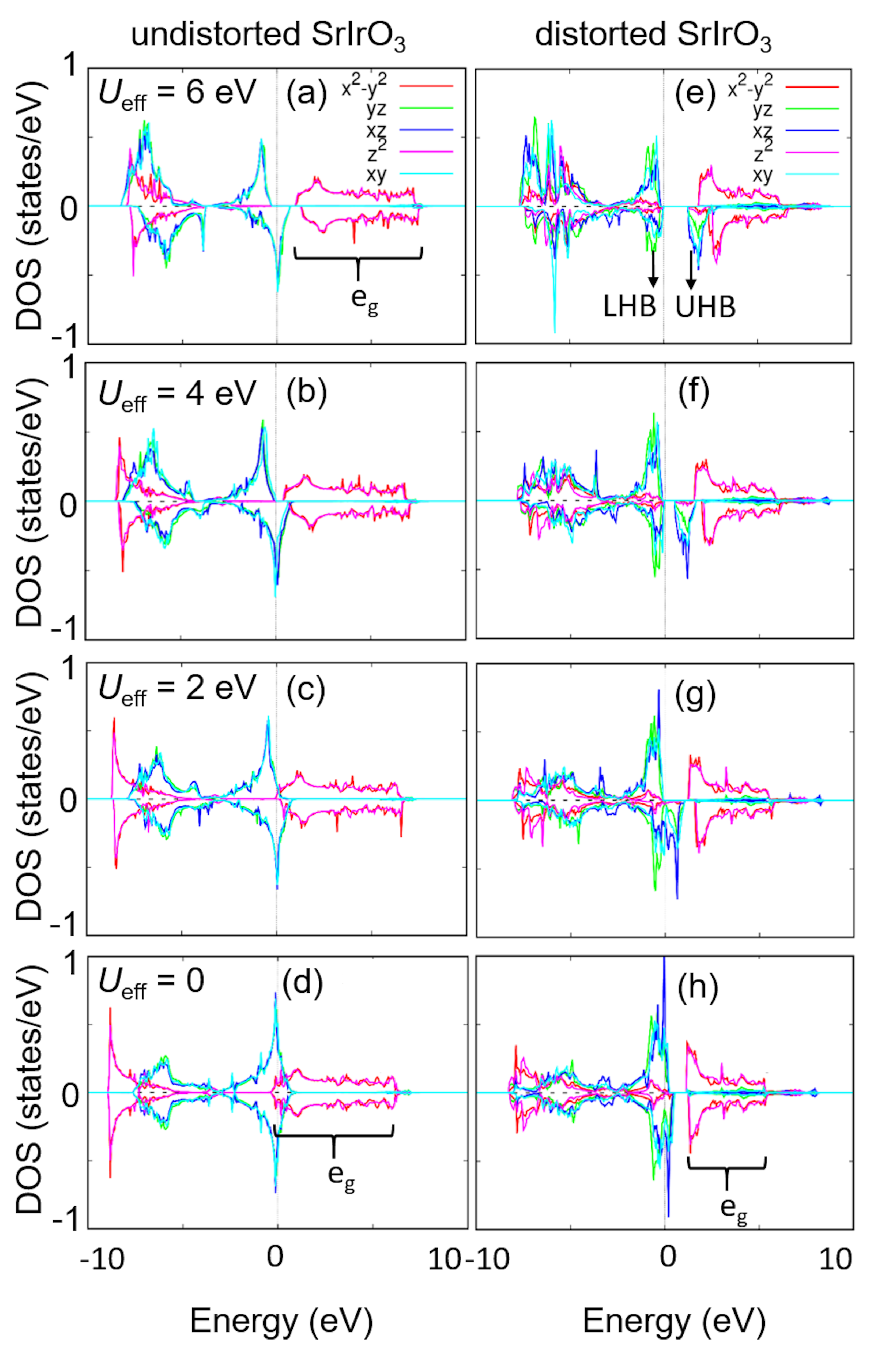}
\caption{Spin and orbital resolved density of states for undistorted (left column) and distorted SIO (right column) for $\lambda$/$\lambda_0$ = 0 and $U_{eff}$ = 0, 2, 4 and 6 eV, respectively.}
\label{Fig13}
\end{figure}

In Fig. \ref{Fig13}, we have plotted the spin and orbital resolved DOS as a function of $U_{eff}$ and for $\lambda$/$\lambda_0$  = 0. As can be clearly seen from the DOS of cubic (undistorted) SIO, for $U_{eff}$ = 0, the e$_g$ states are highly delocalized (bandwidth is $\approx$ 6 eV) and nearly unoccupied.\\
For such states the role of onsite Coulomb repulsion $U_{eff}$ is negligible which indeed is reflected for higher values of $U_{eff}$. In fact, with stronger $U_{eff}$, the $e_{g}$ states are now completely unoccupied. As a result the system will always stabilize in a low-spin (t$_{2g}^5$e${_g}^0$) metallic state. When we introduce the experimentally observed distortions, reduced hopping decreases the bandwidth (see column 2 of Fig. 13). However, the e$_{g}$ bandwidth is still large and there is a distortion induced bandgap which keeps the e$_g$ states unoccupied. In this case increasing $U_{eff}$ creates lower and upper Hubbard bands (LHB and UHB) out of the t$_{2g}$ states and a new band gap emerges at the Fermi level (see Figs. 13 (f,e)).

\bibliography{paper}

\end{document}